\documentclass[fleqn,usenatbib]{mnras} 
\usepackage{mathptmx}
\usepackage[T1]{fontenc}
\usepackage{ae,aecompl}
\usepackage{times}
\usepackage{amssymb}
\usepackage{amsmath}
\usepackage{upgreek}
\usepackage{color}
\usepackage{graphicx}

\usepackage{lscape}
\usepackage{float}
\usepackage{longtable} 

\usepackage{enumitem} 
\setlist{itemsep=4pt}

\title[Evolution of Star Formation in the UDS between z\,=\,\rm{1.46}
and z\,=\,\rm{0.63}]{Evolution of Star
  Formation in the UKIDSS Ultra Deep Survey Field -
  \MakeUppercase{\romannumeral 2}. Star Formation as a
  Function of Stellar Mass Between z\,=\,{1.46} and z\,=\,{0.63}}
\author[Alyssa B. Drake et al.]{Alyssa B. Drake$^{1,2}$, Chris Simpson$^{1}$, Ivan K. Baldry$^{1}$, Phil A. James$^{1}$, Chris A. Collins$^{1}$,\newauthor Masami Ouchi$^{3}$, Suraphong Yuma$^{4,3}$,  James S. Dunlop$^{5}$, Daniel J. B. Smith$^{6}$
\\$^{1}$Astrophysics Research Institute, Liverpool John
 Moores University, IC2, Liverpool Science Park, 146 Brownlow Hill,
 Liverpool L3 5RF, UK
\\$^{2}$CRAL, Observatoire de
Lyon, Universit\'{e} Lyon 1, 9 avenue Charles Andr\'{e}, 69561 Saint Genis
Laval Cedex, France
\\$^{3}$Institute for Cosmic Ray Research, The University of Tokyo,
Kashiwa 277 8582, Japan
\\$^{4}$Department of Physics, Faculty of Science, Mahidol University,
Bangkok 10400, Thailand
\\$^{5}$SUPA, Institute for Astronomy, University of Edinburgh, Royal
Observatory, Edinburgh, EH9 3HJ, UK
\\$^{6}$Centre for Astrophysics, Science and Technology Research
Institute, University of Hertfordshire, Hatfield, Herts, AL10 9AB, UK}

\begin{document}               

\pagerange{\pageref{firstpage}--\pageref{lastpage}} \pubyear{2015}

\maketitle

\label{firstpage}

\begin{abstract} 
We present new results on the evolution of the cosmic star formation
rate as a function of stellar mass in the SXDS-UDS field. We make use of narrow-band selected emission
line galaxies in four redshift slices between $z=1.46$ and $z=0.63$,
and compute stellar masses by fitting a series of templates to
recreate each galaxy's star formation
history. We determine mass-binned luminosity functions in each
redshift slice, and derive the star
formation rate density ($\rho_{\rm{SFR}}$) as a function of mass using
the [O{\sc~iii}] or [O{\sc~ii}] emission lines. We calculate dust
extinction and metallicity as a function
of stellar mass, and investigate the effect of these corrections on the shape of
the overall $\rho_{\rm{SFR}}$(M). We find that both these
corrections are crucial
for determining the shape of the $\rho_{\rm{SFR}}$(M), and its
evolution with redshift. The fully corrected $\rho_{\rm{SFR}}$(M)
is a relatively flat distribution, with the normalisation moving
towards lower values of $\rho_{\rm{SFR}}$ with increasing cosmic
time/decreasing redshift, and requiring star
formation to be truncated across all masses studied here. The peak of $\rho_{\rm{SFR}}$(M) is found
in the $10^{10.5}$\textless$\rm{M_{\odot}}$\textless $10^{11.0}$ mass
bin at $z=1.46$. In the lower redshift slices the location of the peak
is less certain, however low mass galaxies in the range $10^{7.0}$\textless$\rm{M_{\odot}}$\textless $10^{8.0}$ play an important part in the overall
  $\rho_{\rm{SFR}}$(M) out to at least $z\sim1.2$. 
\end{abstract}


\section{Introduction}
\label{intro}
The evolution of cosmic star formation is well-studied, and
crucial to our understanding of the Universe as a whole. It is now
well-established that the peak of
star formation activity lies beyond
$z\sim1$  (e.g. \citealt{Hopkins06}) and that there has been a steep
decline in this activity to the present day (\citealt{Lilly96}, \citealt{LeFloch05},
\citealt{Prescott09}, \citealt{Sobral13},
\citealt{Drake13}). Determining the cosmic
star formation rate density as a function of stellar mass however, $\rho_{\rm{SFR}}$(M), is somewhat
more difficult, and the specifics of this aspect of
observational cosmology are still obscure. Reports of
a shift in the masses of star-forming
galaxies across redshift are now commonplace, however
the meaning of this is complex. Stars in massive galaxies
    are known to have formed
earlier and on a shorter timescale than those formed in lower mass
systems (\citealt{Stanford98}, \citealt{Heavens04},
\citealt{Panter07}), and while this switch-off of star formation in massive galaxies is
readily observed (e.g. \citealt{Brodwin13}) and attributed to the feedback of an active galactic
nucleus (AGN) truncating star formation, the behaviour of the lower
mass star-forming population is less clear. For instance, the
shift of the primary location of star formation activity from higher
to lower mass galaxies with increasing time is clearly seen when
considering specific star formation rates (sSFRs; star formation
rates per unit stellar mass e.g. \citealt{Noeske07}), however, this
places only a weak
constraint on the masses of objects dominating the
overall $\rho_{\rm{SFR}}$ at each $z$. 

In addition to AGN feedback
in high mass galaxies, supernovae are known to expel gas from lower
mass systems preventing the continuation of star formation \citep{Oppenheimer08}, and recent
work has suggested that $\sim 3$ percent of star forming galaxies at
$z \sim 1$ may have their star formation truncated through outflows
of gas (e.g. \citealt{Yuma13}). Determining the role of low-mass
galaxies and establishing the true
shape of the overall $\rho_{\rm{SFR}}$(M) is therefore the first step towards
uncovering the relative importance of the physical processes at work in
truncating star formation.

The Redshift
One LDSS3 Emission Line Survey (ROLES; \citealt{Davies09}) was established to address the
role of low-mass galaxies in the overall  $\rho_{\rm{SFR}}$ as a
function of mass and selected a
sample of dwarf galaxies $10^{8.5}$\textless$\rm{M_{\odot}}$\textless $10^{9.5}$ at $z \sim 1$. 
\cite{Gilbank10} used [O{\sc~ii}] emission to estimate star formation
rates (SFRs) for these
galaxies and concluded that the shape of $\rho_{\rm{SFR}}$(M)
does not evolve with redshift between $z\sim1$ and the present day, a
result since corroborated by \cite{Peng10} and \cite{Sobral14} who
have both reported little change in the mass function of star-forming
galaxies since $z\sim1$. 

The strong [O{\sc~ii}] emission line doublet ($3726$\AA\, and
$3729$\AA) used in \cite{Gilbank10} provides a particularly useful
tool to trace star formation at redshifts \textgreater 1 where
H$\alpha$ is shifted out of the optical window, and yet constraints on
SFR via [O{\sc~ii}] emission are estimated with a number of
caveats. The traditional conversion between observed L[O{\sc~ii}] and SFR
\citep{Kennicutt98} determines the average [O{\sc~ii}]/H$\alpha$ ratio for a sample of 90 local star-forming
galaxies, and uses this to convert the apparent [O{\sc~ii}] luminosity
to an H$\alpha$-derived estimate of the SFR. Applying this relation to
different samples of galaxies however, particularly if these are split by
stellar mass, renders the relationship invalid as the average reddening from H$\alpha$
to [O{\sc~ii}] of samples at different masses may be drastically
different to local star-forming galaxies.

Extinction is a frequently occurring obstacle for extragalactic
astronomy, and many studies have attempted to
quantify the effects and dependence of extinction on
other physical parameters, e.g. stellar mass, metallicity or
SFR (\citealt{Heckman98}, \citealt{Hopkins01}, \citealt{Brinchmann04},
\citealt{Asari07}, \citealt{Garn10}). Using the Balmer decrement,
\cite{Garn&Best10} found that for a star-forming
galaxy, extinction correlates most
strongly with stellar mass, and propose a new relationship to describe
this relationship calibrated for H$\alpha$ luminosity but which can be
extrapolated to other wavelengths. 

As an additional consideration when using a
metal line to determine SFR, one must account for the effect of the
gas-phase metallicity on the strength of the lines. The [O{\sc~iii}]
and [O{\sc~ii}] forbidden lines for instance are sensitive to
metal abundances in addition to the temperature of the gas
(\citealt{Jansen01}, \citealt{Charlot02}). In previous work
\cite{Gilbank10} for example applied an empirical correction to
simultaneously account for the effects of both extinction and
metallicity on the [O{\sc~ii}] indicator. As an alternative however, \cite*{KewleyGeller&Jansen04} determine a conversion
between intrinsic L[O{\sc~ii}] and SFR which is thus independent of the reddening
between the wavelengths of H$\alpha$ and [O{\sc~ii}], and also
includes an optional 
correction for the metallicity of the gas. This relationship can then
be applied to the intrinsic [O{\sc~ii}] luminosities (i.e. those corrected
for extinction at the wavelength of [O{\sc~ii}]) of any sample of galaxies.

Understanding the evolution of $\rho_{\rm{SFR}}$(M) requires a dataset with a number of attributes. In
addition to the requirement of a reliable SFR indicator, the
data must be deep to probe low SFR objects, as well as sampling a wide
range of masses. This problem is
well suited to a narrow-band-selected sample where objects are detected
via line emission across a wide range of masses and are close to
SFR-limited (e.g. \cite{Sobral14} who select star-forming galaxies based on
their H$\alpha$ emission to examine the evolution of the SFR and mass
functions between $z=2.23$ and $z=0.40$).

In this paper, we utilise the ability of narrow-band selection to
detect large numbers of star--forming galaxies irrespective of their
stellar mass, to allow a statistical
evaluation of the objects' properties as a function of mass. We build on
the work of \cite{Drake13} by determining stellar masses for
objects in the four redshift slices covered by deep Subaru data at
$z=0.63$, $z=0.83$, $z=1.19$ and $z=1.46$. We fit maximum likelihood luminosity
functions to our sample using [O{\sc~iii}] or [O{\sc~ii}] emission in
a series of mass bins to examine the
resultant values of $\rho_{\rm{SFR}}$. We incorporate a careful
treatment of extinction and metallicity effects as a function of
stellar mass, calibrating the dependence of SFR on metallicity down to $10^{7.0} < \rm{M_{\odot}} < 10^{8.0}$. 
We aim to highlight the effect of these corrections on the shape of the
$\rho_{\rm{SFR}}$(M) and consider the implications for the physical
processes required to produce the fully corrected
$\rho_{\rm{SFR}}$(M).

This paper
proceeds as follows: in Section 2 we describe in brief the data used
for this analysis, in Section 3 we explain the methods used to
determine stellar masses, produce mass-binned luminosity functions,
calculate star formation rates and the corrections applied to allow a treatment of extinction and metallicity as a function of
stellar mass. In Section 4 we present our results in the form of
luminosity functions and values of $\rho_{\rm{SFR}}$ per mass bin
before discussing the factors affecting our results and the scientific
implications of the shape of the fully corrected
$\rho_{\rm{SFR}}$(M) between redshifts $1.46$ and $0.63$. We then
draw conclusions in Section 6.

For consistency with the manner in which we calculate stellar masses
all values of $\rho_{\rm{SFR}}$ are calculated for a Chabrier IMF. An H$_0$ = 70 kms$^{-1}$ Mpc$^{-1}$, $\Omega_M$ =0.3 and
$\Omega_{\Lambda}$ =0.7 cosmology is assumed throughout, and  all
magnitudes are in the AB system.

\section{Data and Sample Selection}
\label{data}
We use data from the Subaru/XMM-Newton Deep Survey
(SXDS; \citealt{Furusawa08}) and the UKIDSS Ultra Deep Survey (UDS; \citealt{Foucaud07}). The samples discussed
in this paper are drawn from the catalogue of \cite{Drake13}, and are
selected using two narrow-band filters on the Subaru
Telescope at $8150$\,\AA\ (NB816) and $9183$\,\AA\ (NB921). The observations are
complemented by 11 bands of photometry: CFHT $u$-band,
Subaru \emph{B, V, R, i} and \emph{z} bands, UKIRT \emph{J, H} and
\emph{K} bands, and
\emph{Spitzer} \textsc{IRAC} coverage in channels 1 and 2 (see table 1
of \citealt{Drake13} for further details). The narrow-band
    imaging has been smoothed using a Gaussian kernel, in order to match the
    point spread function (PSF) of the SXDS broad-band data
    (\citealt{Ouchi08}, 2009).

Objects are selected according to their narrow-band excess, requiring a $5\sigma$ detection
in the narrow band, a colour excess of at least $3\sigma$ above the
sigma-clipped median intrinsic colour, \emph{and} of at least $3\sigma$ relative
to scatter introduced through photometric uncertainty. 

The sample of line-emitters is cleaned of late-type stellar
contaminants via the \emph{BzK} technique of \cite{Daddi04}, and
photometric redshifts are determined using the photometric
redshift-fitting software ``EA$z$Y'', making full use of the 11-band
photometry to assign objects to redshift slices. This approach
    results in photometric redshifts with a normalised median absoulte
    deviation of $\sigma$$_{\rm{NMAD}}= 0.026$.

Finally, the completeness of detection in the narrow-band sample is assessed per
    $\Delta\rm{{m}_{NB}}=0.05$ bin using randomly positioned fake
    objects. The stringent $5\sigma$ detection limit means the effect
    is small, and only small numbers of objects need to be accounted
    for in the maximum likelihood analysis.

Full details can be found in \cite{Drake13} of
data coverage, sample selection, photometric redshift determination
and assignment to redshift slices (via stacked probability density
distributions) plus further information on the maximum likelihood
approach to luminosity functions. Table
\ref{tab:no.s} gives numbers of objects in each redshift slice used
here. 

\begin{table}
\caption[]{Objects per redshift slice}
\label{tab:no.s}
\begin{center}
\begin{tabular}{cccc} \hline
Filter & Redshift slice & Emission Line & Objects\\
\hline
NB816 & 0.35 $<$ {\bf{0.63}} $<$ 0.80 & [O{\sc~iii}] & 999 \\ 
NB921 & 0.50 $<$ {\bf{0.83}} $<$ 1.10 & [O{\sc~iii}] & 894 \\ 
NB816 & 0.80 $<$ {\bf{1.19}} $<$ 1.50 & [O{\sc~ii}] & 956 \\ 
NB921 & 1.10 $<$ {\bf{1.46}} $<$ 1.90 & [O{\sc~ii}] & 2203 \\ 
\hline
\end{tabular}
\end{center}
\end{table} 

\section{Method}
\subsection{Stellar Masses}
We determine stellar masses following the method of
\cite{Simpson13}. Each object's redshift is set to the median
    redshift of the redshift slice, and the SED is fit with a series
of synthetic star formation histories (SFHs) using the redshift--fitting sofware EA$z$Y to determine the most probable scenario for the
galaxy's assembly. \cite{Simpson13} design a
set of 40 SFH templates based on instantaneous starbursts using a
    Chabrier IMF and the spectral models of Charlot and Bruzual (2007). The SFH
templates range between very young populations, $\sim20 \times 10^6$
years old, up to populations $13 \times 10^9$ years old. Finer time resolution is
used between the younger templates, and a few young reddened templates
are also included. The advantage of this approach is the ability to
recreate \emph{any} SFH through combination
of instantaneous starburst templates. As this calculation is based on
2 arcsecond aperture fluxes from the catalogue described in
\cite{Drake13} however, each mass must be scaled by the
ratio of total K-band flux to aperture K-band flux. Here we use total K flux
values from the catalogue of \cite{Grutzbauch11} where available, or a
2.2 arcsecond aperture flux to simulate total K-band flux otherwise.

\begin{figure}
\begin{center}
\resizebox{0.48\textwidth}{!}{\includegraphics{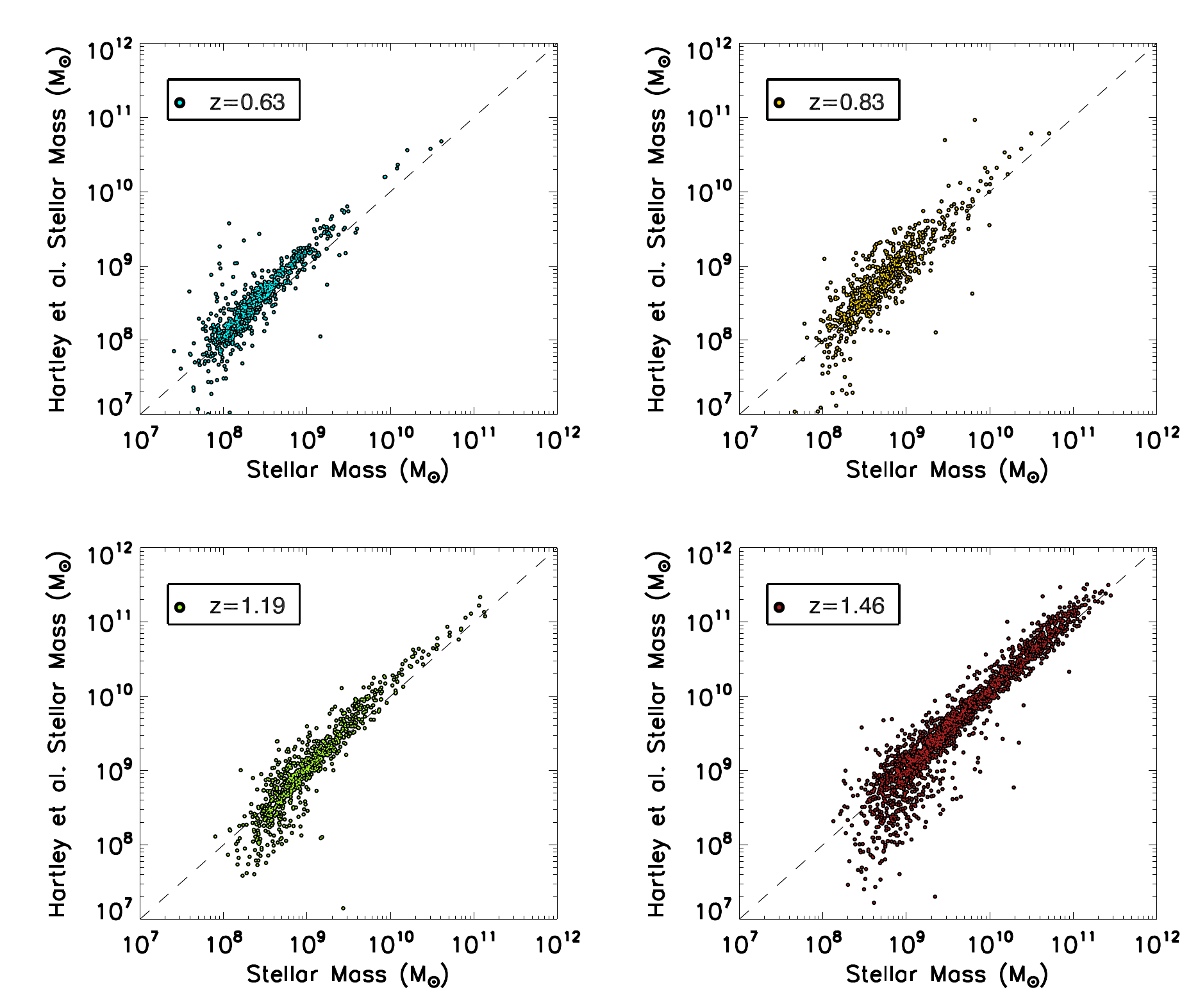}} 
\caption[Stellar mass estimates from this work in comparison to
Hartley et al. 2013]{Comparison of stellar mass estimates from this work and
  results in the literature from Hartley et al. 2013. The four panels
  represent the 4 redshift slices we have used for this analysis
  progressing from top left to bottom right in increasing redshift order $z=0.63, 0.83,
  1.19, 1.46$.}
\label{fig:Will_mass}
\end{center}
\end{figure}

Figure \ref{fig:Will_mass} shows a
comparison of masses derived in this work and the stellar mass
estimates of the same objects detected in
\cite{Hartley13}. Not all of the star-forming galaxies detected in our
survey are found in the K-band selected sample of \cite{Hartley13}, however
where the samples overlap our stellar masses agree well with
the published work. Our use of the
Charlot and Bruzual (2007) models which incorporate a greater contribution
from thermally pulsing (TP-) AGB stars, results in lower masses by
$0.1-0.2$ dex than those of the \citealt{Bruzual&Charlot03} models used in
\cite{Hartley13}.

\subsection{Mass-Binned Maximum-Likelihood Luminosity Functions}
\label{ML}

\begin{figure}
\begin{center}
\resizebox{0.48\textwidth}{!}{\includegraphics{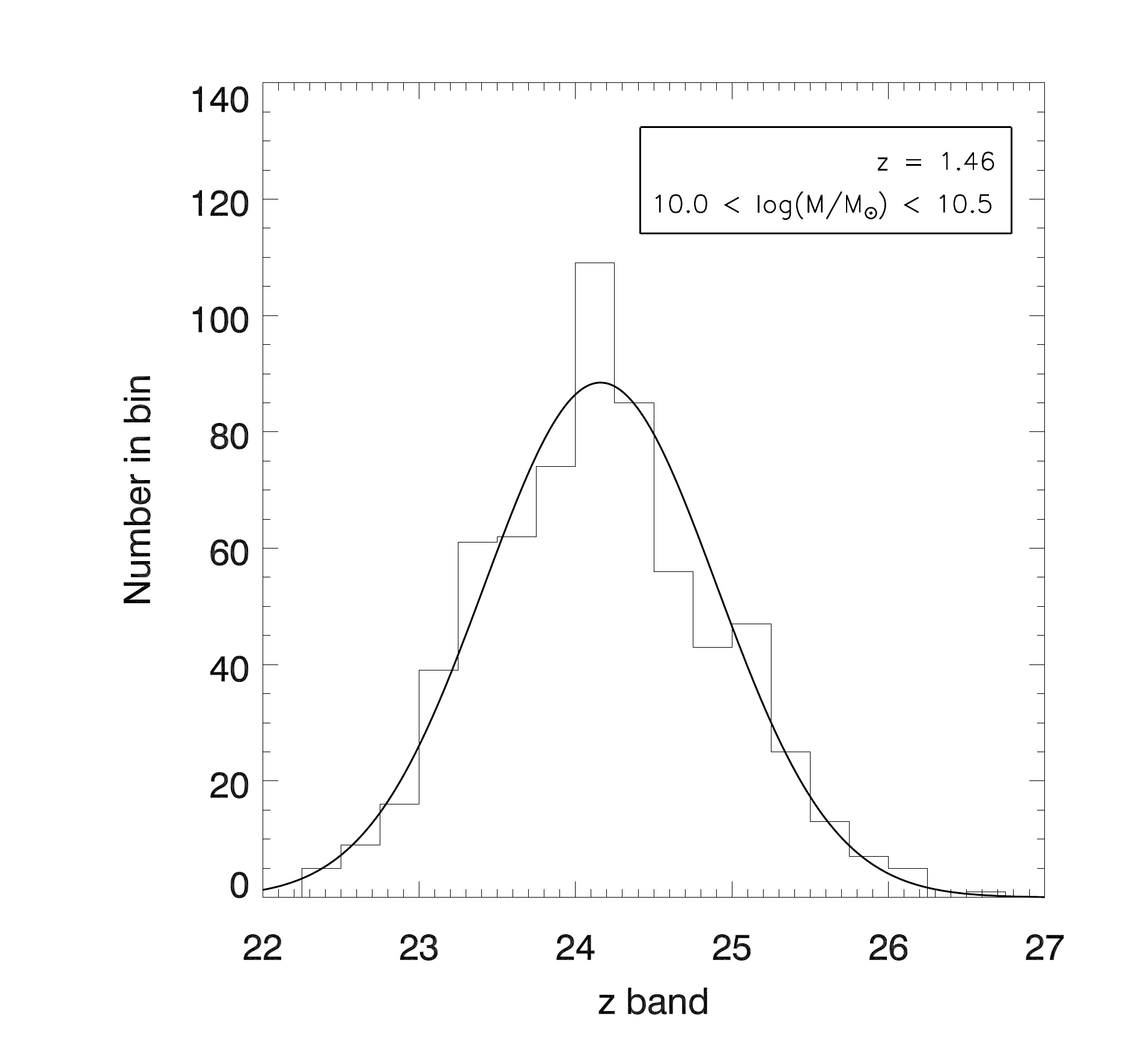}} 
\caption[Gaussian fit to distribution of broad-band magnitudes in a
particular mass bin]{Example of Gaussian fit to the distribution of
  broad-band magnitudes at $z=1.46$ in the $10.0$ \textless\
  log(M/M$_{\odot}$) \textless\ $10.5$ mass bin. For these
  NB921-selected objects this is the z band magnitude, taken from the
  K-band selected catalogue of \cite{Simpson13}}
\label{fig:Gaussian_fit}
\end{center}
\end{figure}

\cite{Drake13} described our method of determining maximum
likelihood luminosity functions for narrow-band selected samples, and demonstrated that this approach possesses a number of
advantages over existing techniques. We now apply this method to
mass-binned data at $z=0.63$, $z=0.83$, $z=1.19$
and $z=1.46$ to determine luminosity functions as a function of
stellar mass. A small but important difference between this analysis and that of \cite{Drake13}, 
is the manner in which we model the underlying star-forming galaxy
population. For the original analysis the population of broad-band
selected galaxies in a redshift slice was well fit by a Schechter
function, however splitting the sample by
stellar mass, this approach is obviously no longer appropriate. We find the best
fit when this population is modelled as a Gaussian (see Figure
\ref{fig:Gaussian_fit}), and so we use this to represent the broad-band
magnitudes of galaxies in each mass
range from the K-band selected catalogue of
\cite{Simpson13}.  For each redshift slice we fit to the
observed broad band corresponding to the rest-frame wavelength of the
NB selection filter, i.e. for NB816-selected objects this is an
interpolation of the i and z bands, and for NB921-selected objects
z-band only. The \cite{Simpson13} catalogue is used here in preference to
\cite{Grutzbauch11} or \cite{Hartley13} for consistency in stellar
mass estimates only.  

Each redshift slice is split
into $5-7$ mass bins (depending on the number of detections and range
of stellar masses) of width
$1$ dex in the lowest mass bin ($10^{7.0}$\textless$\rm{M_{\odot}}$\textless $10^{8.0}$) and $0.5$ dex for the remainder of
the sample. 

\subsection{Star Formation Rates}
To investigate the effects of different assumptions about dust and
metallicity on the resultant SFRs, we compute $\rho_{\rm{SFR}}$
using a series of different approaches.\\ 
To determine SFR from L$_{\rm{[OIII]}}$ we initially take the same approach as in
\cite{Drake13} and apply the standard line ratios:
H$\alpha$/H$\beta$=2.78 and [O{\sc~iii}]/H$\beta$=3
\citep{Osterbrock&Ferland89} to convert the \cite{Kennicutt98}
relation between L$_{\rm{{H\alpha}}}$ and SFR to one for L$_{\rm{[OIII]}}$ (Equation \ref{eq:SFROIII}):

\begin{equation}
{{\rm{SFR}}({\rm{M}}_{\odot}{\rm{yr}}^{-1}) = 7.35 \times
  10^{-35}{{\rm{L_{[OIII]}}}} {{\rm{E}_{[OIII]}}}} 
\label{eq:SFROIII}
\end{equation}

\noindent where L$_{\rm{[OIII]}}$ is the observed [O{\sc~iii}] luminosity,
and E$_{\rm{[OIII]}}$ represents extinction at the
wavelength of [O{\sc~iii}]. The \cite{Kennicutt98} approximation of
SFR for L$_{\rm{[OII]}}$ is computed according to
Equation \ref{eq:SFROII}:

\begin{equation}
{{\rm{SFR}}({\rm{M}}_{\odot}{\rm{yr}}^{-1}) = 1.39 \times
  10^{-34}{{\rm{L_{[OII]}}}} {{\rm{E}_{H \alpha}}}}
\label{eq:SFROII}
\end{equation}

\noindent where L$_{\rm{[OII]}}$ is the observed [O{\sc~ii}] luminosity,
and E$_{\rm{{H\alpha}}}$ represents extinction at the
wavelength of H$\alpha$.

\subsubsection{Mass- Dependent Extinction Correction}
Extinction is known to correlate with a number of physical
    properties of galaxies (\citealt{Brinchmann04}, \citealt{Garn09b}) the dominant of these being stellar
    mass \citep{Garn&Best10}.
To address the mass-dependence of dust extinction, we implement Equation
\ref{eq:Garn_A_Halpha} of \cite{Garn&Best10} to describe H$\alpha$
extinction in magnitudes as a
function of stellar mass: 

\begin{equation}
{A_{\rm{{H\alpha}}} = 0.91+0.77x+0.11x^2-0.09x^3}
\label{eq:Garn_A_Halpha}
\end{equation}

\noindent where $x={\rm{log_{10}}(M/10^{10}M_{\odot})}$. Using Equation
\ref{eq:Garn_A_Halpha} we determine values for A$_{\rm{H\alpha}}$ per
mass bin and apply the \cite{CardelliClaytonMathis}
reddening law to the set of values found for this prescription to determine the corresponding values of A$_{\rm{{[OIII]}}}$
and A$_{\rm{{[OII]}}}$. Each fit is set to a
constant below $10^{9.0}{\rm{M_{\odot}}}$ where the estimate is unreliable.

For L$_{\rm{[OIII]}}$ estimates of SFR we use Equation
\ref{eq:SFROIII} in conjunction with an assessment of the dust
extinction per mass bin from Equation \ref{eq:Garn_A_Halpha}. For
L$_{\rm{[OII]}}$ however, the estimate of SFR from Equation \ref{eq:SFROII}
    is affected by uncertainties due to the wavelength dependence of
    reddening between the H$\alpha$ and ${\rm{[OII]}}$ lines. While this is the standard approach for
    large statistical samples, it is unsuitable for this dataset, binned by stellar mass and
    spanning a wide range of redshifts. \cite{KewleyGeller&Jansen04} re-calibrate the
    ${\rm{[OII]}}$ indicator to be reddening independent, allowing for
    the application of SFR(${\rm{[OII]}}$) to a wide range of galaxy
    samples. This is presented in Equation \ref{eq:[OII]_conversion}:

\begin{equation}
{{\rm{SFR([OII])(M_\odot yr^{-1})}} = (6.58  \pm 1.65) \times 10^{-35}
  {{\rm{L_{[OII]}}}}  {{\rm{E}_{[OII]}}}}
\label{eq:[OII]_conversion}
\end{equation}

\noindent where L$_{\rm{[OII]}}$ is observed [O{\sc~ii}]
luminosity, and E$_{\rm{[OII]}}$ is the flux extinction at the
wavelength of [O{\sc~ii}]. We apply this
conversion to
determine $\rho_{\rm{SFR}}$ for galaxies with a dust extinction correction
as a function of stellar mass, using Equation
    \ref{eq:Garn_A_Halpha} in conjunction with the prescription of
    \cite{CardelliClaytonMathis} to determine E$_{\rm{[OII]}}$.

\subsubsection{Metallicity Correction}

Although the [O{\sc~iii}] and [O{\sc~ii}] forbidden lines are
sensitive to the the metallicity of a galaxy, Figure 11 of \cite{KewleyGeller&Jansen04} shows that the ratio of
[O{\sc~iii}]/H$\beta$ line luminosities is constant down to very low
metallicities for a range of ionisation states of the gas, and hence we apply no metallicity
correction for [O{\sc~iii}]-derived estimates of SFR.

[O{\sc~ii}] emission however varies
considerably with gas-phase metallicity (here meaning oxygen
abundance defined as \mbox{log(O/H) + 12}), and since the traditional
conversion of L$_{\rm{[OII]}}$ to SFR relies on the relationship between observed [O{\sc~ii}] to
H$\alpha$, the variation with metallicity introduces an error in this approach. \cite{KewleyGeller&Jansen04} 
build on their earlier work to incorporate the effect of metallicity
on the [O{\sc~ii}]/H$\alpha$ ratio, and hence derive a theoretical
calibration for SFR(L$_{\rm{[OII]}}$) as a function of
L$_{\rm{[OII]}}$  and
\mbox{metallicity. \footnote{subscript \emph{t} refers to the fact this is a
  theoretical prediction} :}

\begin{equation}
{{\rm{SFR([OII],Z)_t(M_\odot yr^{-1})}} = \frac{7.9 \times 10^{-35} {\rm{L([OII])}}}{\rm{[OII]/H\alpha}}}
\label{eq:[OII]_mass_Z}
\end{equation}

\begin{figure}
\begin{center}
\resizebox{0.48\textwidth}{!}{\includegraphics{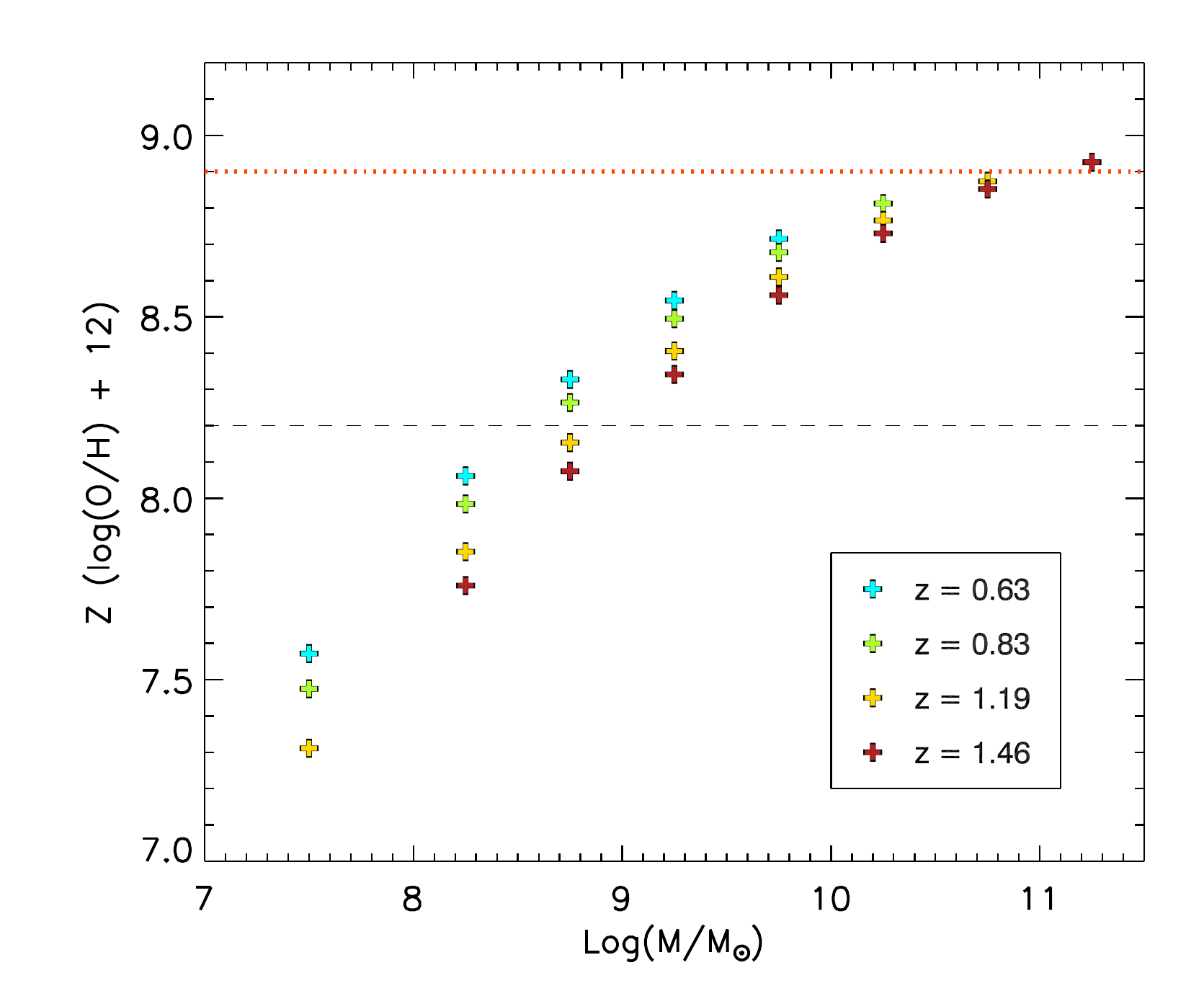}} 
\caption[Metallicity estimate per mass bin]{Metallicity per mass bin of width 0.5 dex. Metallicity is
particularly low in the lower half of mass bins. Solar abundance in
these units is 8.9 marked by the dark orange dotted line. The dashed
horizontal line at 8.4 represents the limit of the metallicities
modelled in the Kewley et al. (2004) calibration for SFR.}
\label{fig:Z}
\end{center}
\end{figure}

They use theoretical models to examine the dependence of the
[O{\sc~ii}]/H$\alpha$ ratio on metallicity for a number of different ionisation
states of the H{\sc~ii} gas. The
relationship for intrinsic [O{\sc~ii}]/H$\alpha$ line ratio on metallicity is
then given by:

\begin{equation}
{{\rm{[OII]/H\alpha}} = a + b{\rm{Z}} + c{\rm{Z^2}} + d{\rm{Z^3}}}
\label{eq:[OII]_mass}
\end{equation}

\noindent where Z= log(O/H) + 12, and and the coefficients $a$, $b$, $c$, and $d$ originate from the best fit
curve for metallicity abundance vs [O{\sc~ii}]/H$\alpha$ for the
appropriate ionisation parameter. 

For consistency with our assumption
of [O{\sc~iii}]/H$\beta$=3 in \cite{Drake13}, we adopt an ionisation
parameter\footnote{where $q$ is the maximum velocity of the
ionisation front driven by the local radiation field
\citep{KewleyGeller&Jansen04}} $q =4 \times 10^7$. This is equivalent to \mbox{log $U = -2.875$}, where $U$,
the commonly adopted ionisation parameter, is defined as
$U=q/c$. Our choice is entirely consistent with recent detailed studies
e.g. \cite{Nakajima14} who found  $q = 1
\times 10^7 - 1 \times 10^8$, and also falls within the ``normal'' range of
ionisation parameters found in \cite{KewleyGeller&Jansen04}, and \cite{Dopita00}: $q = 1
\times 10^7 - 8 \times 10^7$ cm s$^{-1}$.  For $q =4 \times 10^7$, $a=-1432.67$, $b=470.545$,
$c=-51.2139$  and $d=1.84750$.

To derive metallicities for our sample we employ
the empirical model of \cite{Savaglio05} for the evolution of the mass-metallicity relation to high
redshift. They derive an expression for the metallicity
of a galaxy of a given stellar mass, at t$_H$, the age of the Universe at
that redshift for the assumed cosmology, as:

\begin{align}
12 + {\rm{log (O/H)}} = -7.5903 +2.5315\ {\rm{log}}\ M \nonumber \\ 
- 0.09649 \ {\rm{log^2}} M + 5.1733\ {\rm{log}}\ t_{\rm{H}}  -0.3944\ {\rm{log^2}}\ t_{\rm{H}}  \nonumber \\
 - 0.4030\ {\rm{log}}\ t_{\rm{H}}\ {\rm{log}}\ M. \nonumber \\
\label{eq:Z}
\end{align}

We use Equation \ref{eq:Z} to determine an average metallicity per
mass bin, for each redshift slice. The resultant values can be seen in
Figure \ref{fig:Z}, and reach substantially sub-solar values. The metallicity of the
Sun is shown on this plot as the dotted orange line at Z=8.9. 

For mass bins of very low metallicity (\textless\ 8.2, about half of
our sample) the model fit to [O{\sc~ii}]/H$\alpha$ from
\cite{KewleyGeller&Jansen04} breaks down, and extrapolating the fit results in non-physical values of
[O{\sc~ii}]/H$\alpha$. The limit of the \cite{KewleyGeller&Jansen04}
model is shown as the dashed black line in Figure \ref{fig:Z}. \cite{KewleyGeller&Jansen04} use \textsc{Pegase} stellar
population models \citep{PEGASE} with the \textsc{Starburst99} code
\citep{Leitherer99} to determine an ionising
spectrum to 
simulate values of [O{\sc~ii}]/H$\alpha$, using the \textsc{Mappings{\sc~iii}}
radiative transfer code (e.g. \citealt{Sutherland93}) for different values of metallicity and ionisation
parameter, $q$. 

To determine values of [O{\sc~ii}]/H$\alpha$ we therefore follow the method of \cite{KewleyGeller&Jansen04}
but use the \textsc{CLOUDY} (Ferland et al. 2013) radiative transfer code to
determine [O{\sc~ii}]/H$\alpha$ at each metallicity. We assume a total H density
of $10^{2.5}$ cm$^{-3}$ which corresponds to the electron density
$350$ cm$^{-3}$ used in \cite{KewleyGeller&Jansen04}. The resultant 
values of [O{\sc~ii}]/H$\alpha$ agree to $\sim 10 \%$ over the corresponding metallicity range. 

The [O{\sc~ii}]/H$\alpha$ ratios are well fit
in log space by Equation \ref{eq:[OII]_ha_me}:

\begin{equation}
{{\rm{log([OII]/H\alpha})} = a + b{\rm{Z}} + c{\rm{Z^2}} + d{\rm{Z^3}}}
\label{eq:[OII]_ha_me}
\end{equation}

\noindent with best fitting coefficients $a =76.94$, $b=-33.86$, $c=4.77$,
and $d=-0.22$. 

\begin{figure*}
\begin{center}
\resizebox{0.48\textwidth}{!}{\includegraphics{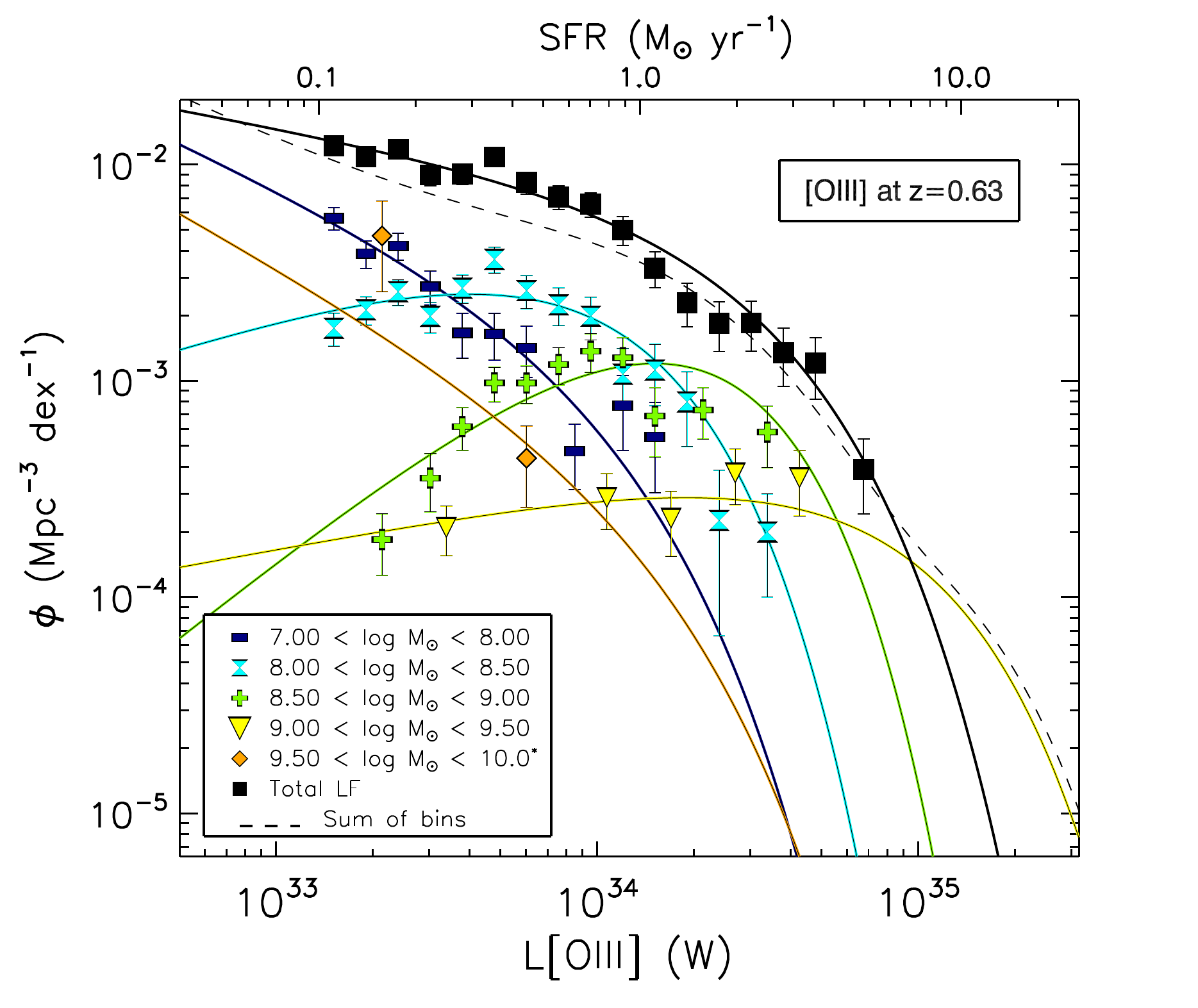}} 
\resizebox{0.48\textwidth}{!}{\includegraphics{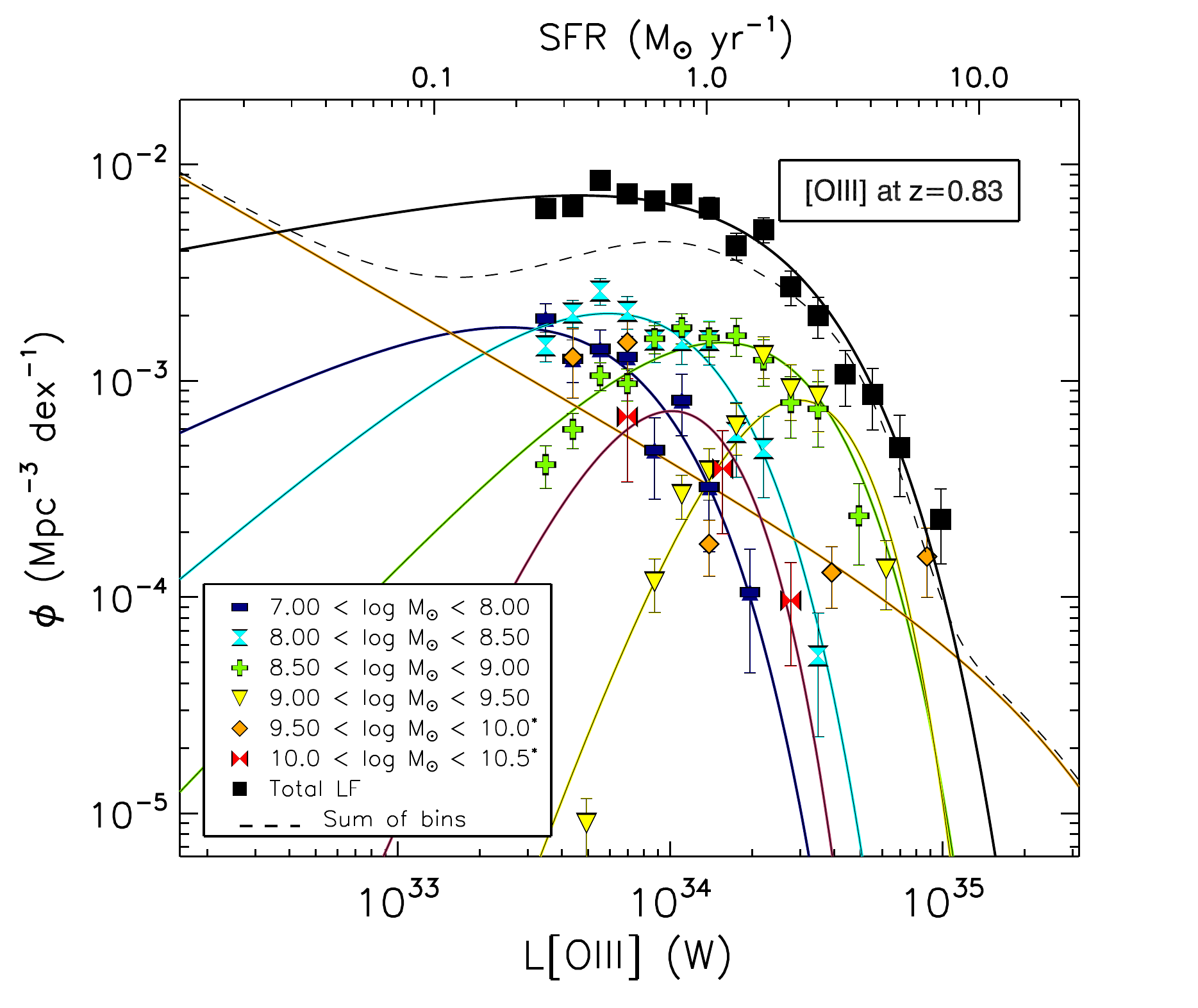}} 
\resizebox{0.48\textwidth}{!}{\includegraphics{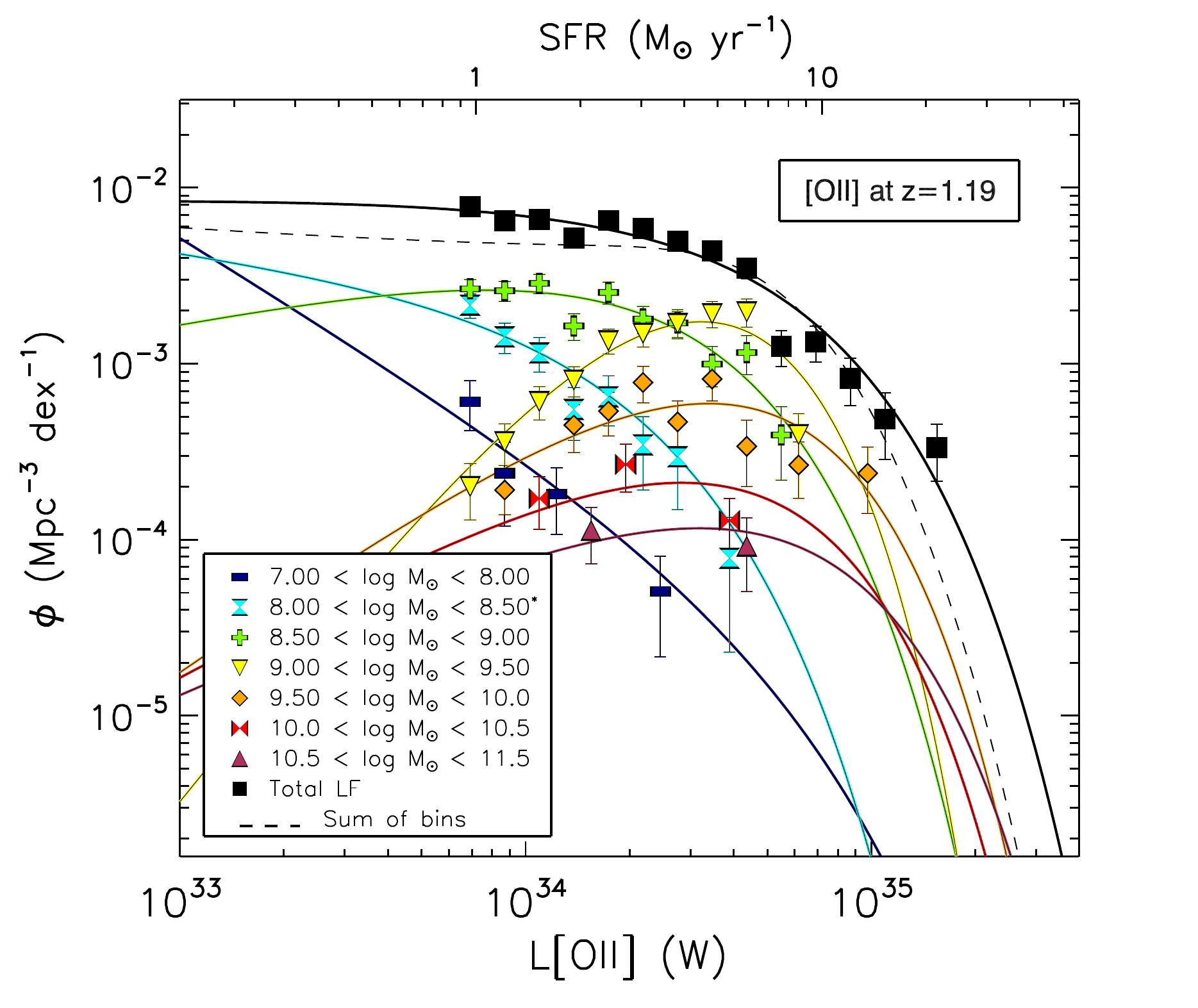}} 
\resizebox{0.48\textwidth}{!}{\includegraphics{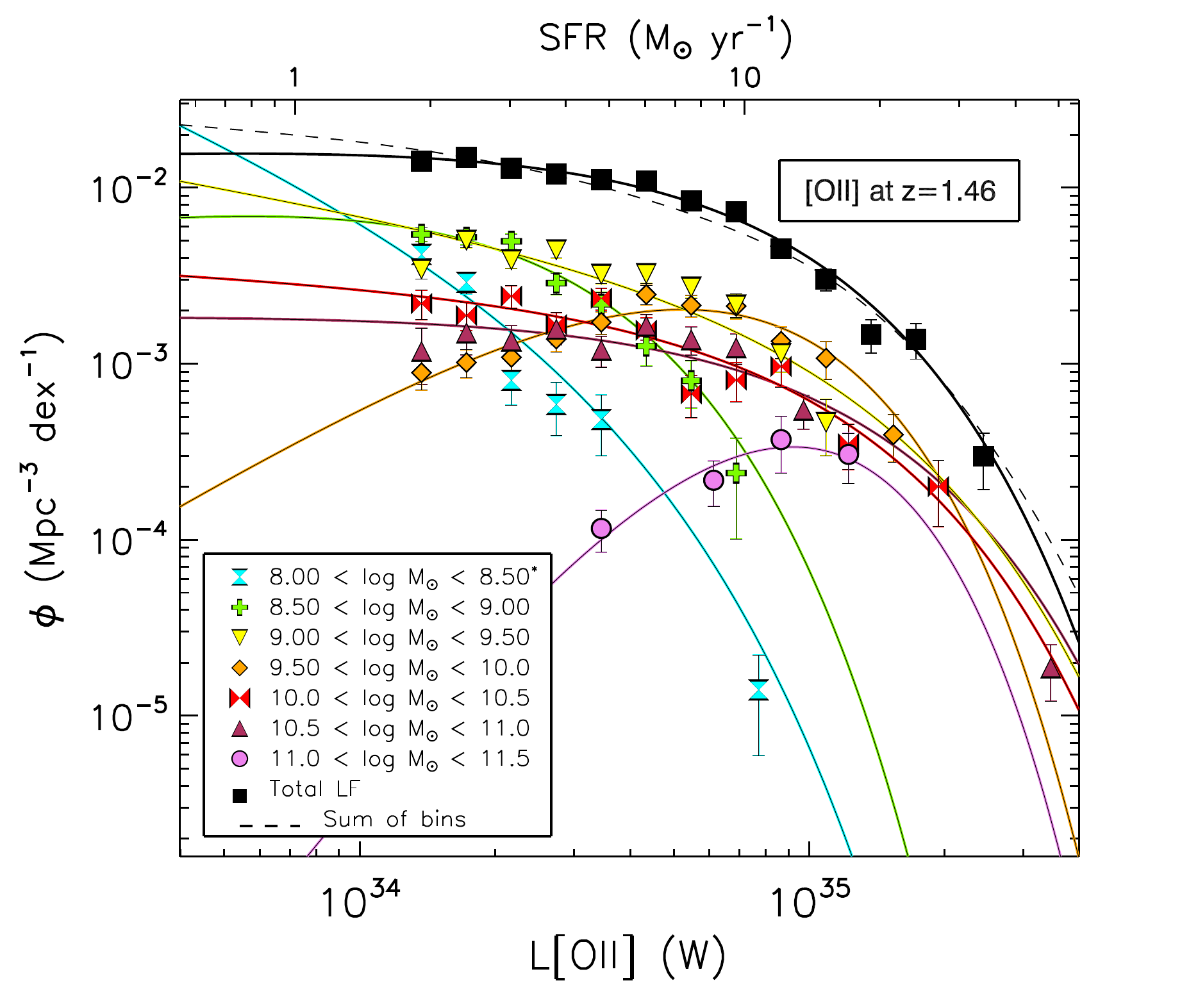}} 
\caption[]{Best--fitting luminosity
  functions for redshift slices: $z=0.63$ between 7.25 and 9.25
  dex, $z=0.83$ ( 7.75 to 10.25 dex),
  $z=1.19$ (7.75 to 10.75 dex), $z=1.46$ (7.75 to 11.25 dex). The
  upper SFR scale on these plots gives an indication of SFR
  values assuming a constant scaling from luminosity for these LFs. Neither these numbers nor the luminosities
  have been corrected for dust
extinction.}
\label{fig:Mass_LFs}
\end{center}
\end{figure*}

We use Equation \ref{eq:[OII]_mass_Z} in conjunction with our own
estimates of [O{\sc~ii}]/H$\alpha$ to determine fully corrected values
of $\rho_{\rm{SFR}}$ based on [O{\sc~ii}] luminosity.


\section{Results}
\label{results}
\subsection{Luminosity Functions}
The mass-binned maximum-likelihood luminosity functions for objects at the
    weighted-mean redshift of emitters in a particular redshift
    slice are presented in Figure
\ref{fig:Mass_LFs} and Table \ref{tab:massLF}. Volumes quoted in the table are representative of the
volume where narrow-band filter transmission is greater than
0.5. Values of L$^*$ are corrected for Galactic extinction and
aperture effects. The data in Figure
\ref{fig:Mass_LFs} are presented as in \cite{Drake13}, where the solid
coloured line gives the maximum likelihood fit, and the data points on
the plot are binned arbitrarily for presentation.

For each luminosity function the maximum likelihood analysis
determines the most likely values of $\phi^*$, L$^*$ and $\alpha$,
allowing all three parameters to vary. In order to produce a
well-constrained fit however, the data must probe significantly below
L$^*$ in the particular redshift/mass-bin combination. For some
luminosity functions, the depth of the data  coupled with a small
number of emitters in that mass bin, result in poorly-constrained values
of $\alpha$. Where this is the case, we follow the approach taken in
\cite{Drake13}, and constrain the luminosity function with a Gaussian
prior on the faint end slope, set to the value of $\alpha$ for the closest
mass bin in that redshift slice. A point of note is that in each case the sum of the mass-binned luminosity
functions (denoted by the dashed line) including those with a Gaussian prior, is in good agreement with the
best--fit luminosity function for the entire redshift slice (given by
the solid black line). The small deviations between the two lines seen
at the bright or faint end of the LFs can be explained by objects in
the wings of the mass distributions where there were too few sources
to fit a LF.

\subsection{$\rho_{\rm{SFR}}$ as a Function of Stellar Mass}
\label{sect:SFRD(mass)}

Figure \ref{fig:SFRD(mass)} presents values of
$\rho_{\rm{SFR}}$(M) evaluated considering an
integration of the data to two different limits in the
left and right hand panels, and
incorporating various different approaches to the treatment of dust
extinction and metallicity. The method used for each row of panels is
detailed below and the corresponding
values are summarised in Table \ref{tab:SFRD}. For consistency with
the manner in which we calculate our stellar masses, the
$\rho_{\rm{SFR}}$ values quoted and plotted have been converted to a Chabrier IMF. 

Initially, $\rho_{\rm{SFR}}$ is computed using the \cite{Kennicutt98}
relation and applying 1 magnitude of extinction at H$\alpha$ (upper
panels Figure \ref{fig:SFRD(mass)}). In the central panels of Figure
\ref{fig:SFRD(mass)}, we determine extinction as a function of
stellar mass according to the prescription of
\cite{Garn&Best10} using \cite{CardelliClaytonMathis} to infer values at [O{\sc~iii}] and
[O{\sc~ii}], and use Equation \ref{eq:[OII]_conversion} of
\cite{KewleyGeller&Jansen04} to
derive SFR([O{\sc~ii}]). Finally, in the lower two panels we use our estimates of
log(O/H) + 12 to incorporate a metallicity correction in conjunction
with Equation \ref{eq:[OII]_mass_Z} of \cite{KewleyGeller&Jansen04} to
determine a fully corrected $\rho_{\rm{SFR}}$(M). 

The appropriate limits of integration to determine
$\rho_{\rm{SFR}}$ per mass bin are complicated for this analysis,
since the sensitivity in $\rm{M_{\odot}} yr^{-1}$ varies considerably
due to the use of two different emission line indicators and the range
of redshift studied. Here we evaluate each luminosity function in two
different ways; to the limit of
the data in each redshift slice (left-hand panels Figure \ref{fig:SFRD(mass)}) and to the same limiting
SFR in each redshift slice (right-hand panels Figure
\ref{fig:SFRD(mass)}). In the former approach, the limiting SFRs are $0.08, 0.12, 0.55$ and $1.42\, \rm{M_{\odot}}
yr^{-1}$ at $z=0.63, 0.83, 1.19$ and $1.46$ respectively, producing sensible results for most mass bins. Intergrating to the same limiting
SFR for comparison across redshift slices however presents more of a
challenge. Selections made in the $z=0.63$ and $z=0.83$ redshift
slices can be integrated to low SFRs $\sim 0.1 \rm{M_{\odot}}
yr^{-1}$, however limits which produce sensible results for these
redshift slices require a large extrapolation at $z=1.19$ and $z=1.46$ and
consequently introduce a far greater uncertainty on luminosity
functions that probe little below L$^*$. Likewise, limiting the
integration to values where the luminosity function is
well-constrained in these redshift slices results in the loss of a
large portion of the $\rho_{\rm{SFR}}$(M) in the lower redshift
slices (additionally there is a dramatic drop in the
$10^{9.5}$\textless$\rm{M_{\odot}}$\textless $10^{10.0}$ bin at
$z=0.63$  where the whole luminosity function is poorly constrained) and ultimately the best choice for a constant limiting SFR is $1 \rm{M_{\odot}}
yr^{-1}$. We include this analysis to provide a consistent evaluation across redshift, however we prefer values of $\rho_{\rm{SFR}}$(M)
based on the limit of the data as being representative of the total SFR.

\subsubsection{Errors on $\rho_{\rm{SFR}}$(M)}
\begin{figure*}
\begin{center}
\resizebox{0.48\textwidth}{!}{\includegraphics{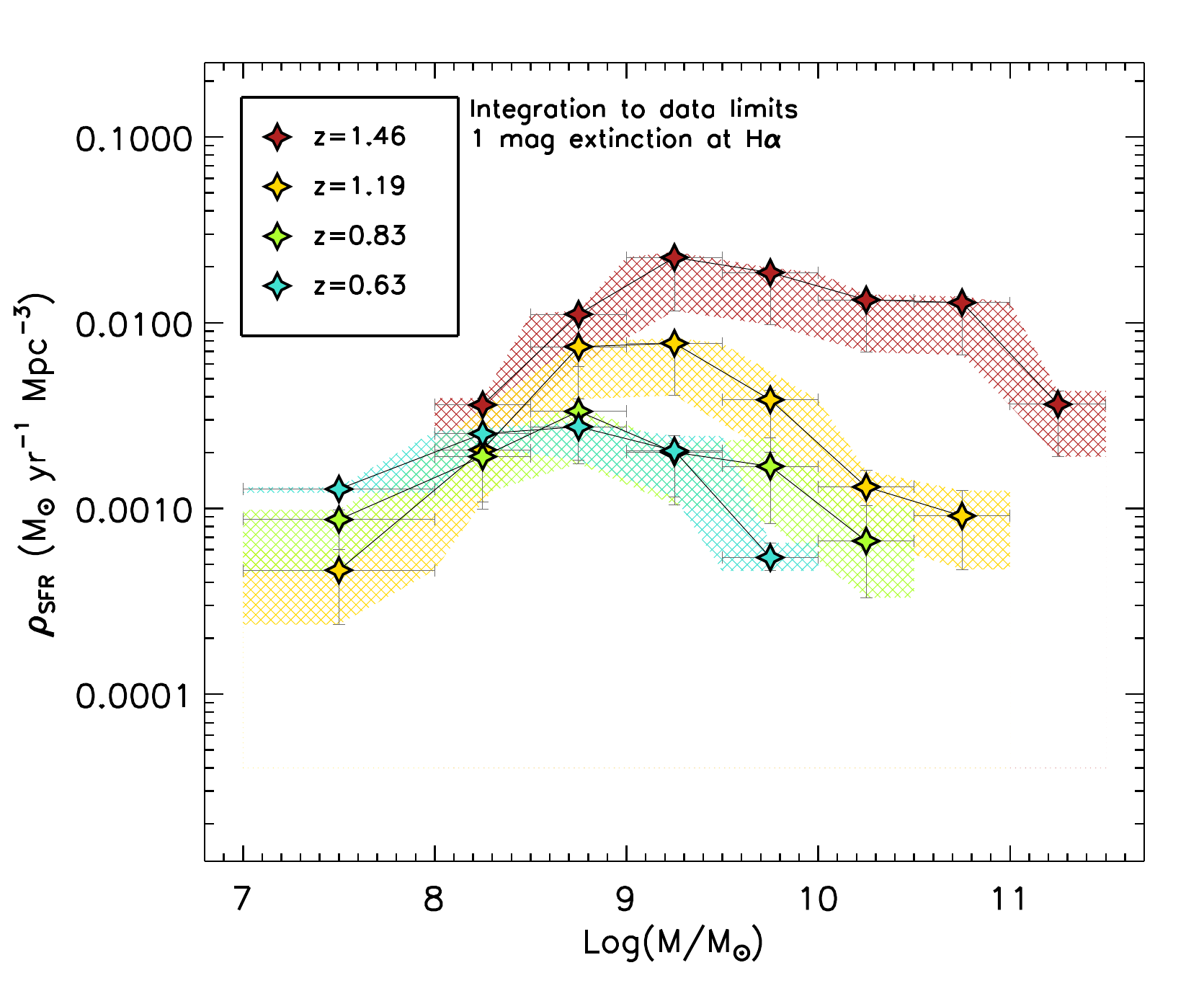}} 
\resizebox{0.48\textwidth}{!}{\includegraphics{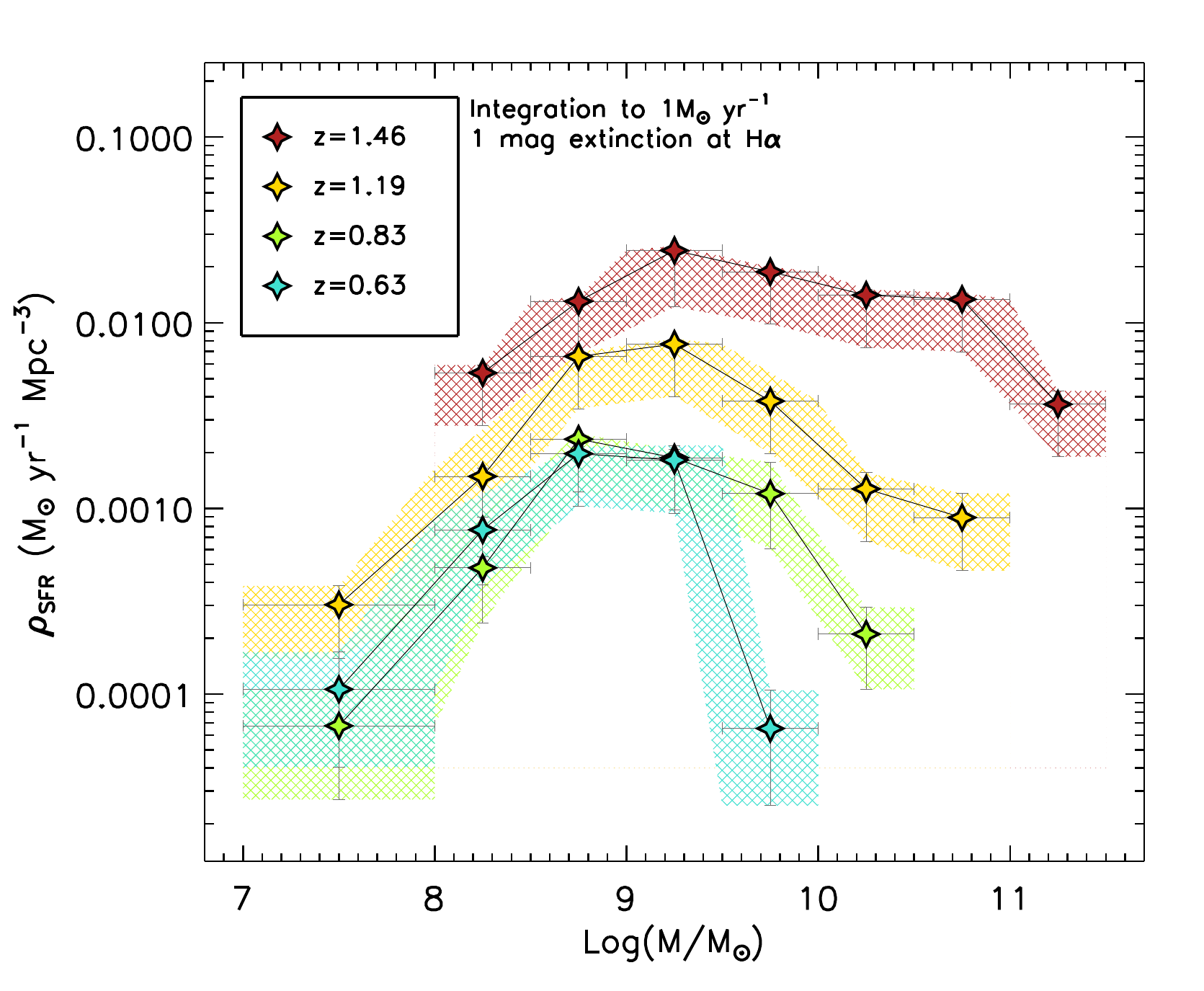}} 
\resizebox{0.48\textwidth}{!}{\includegraphics{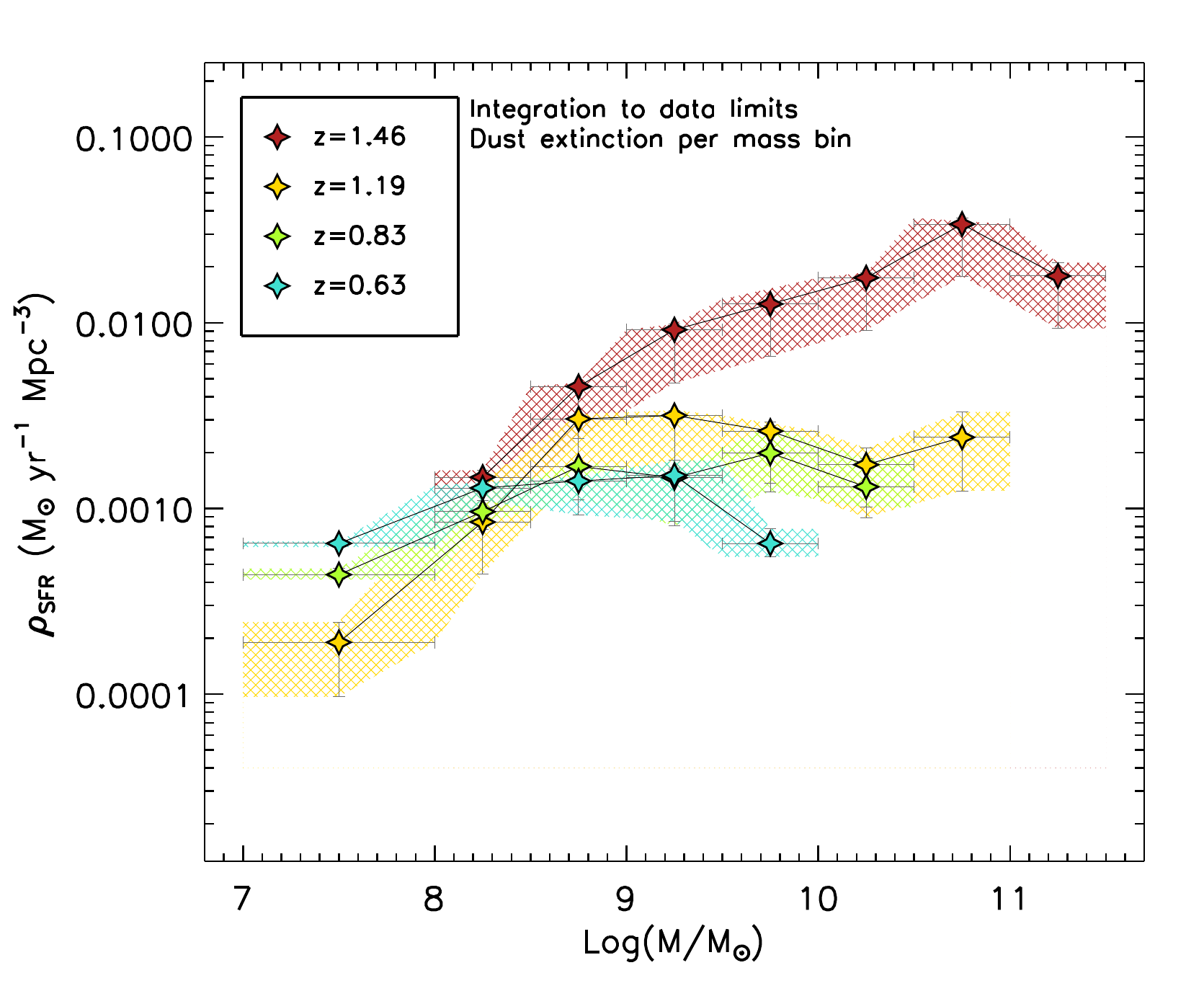}} 
\resizebox{0.48\textwidth}{!}{\includegraphics{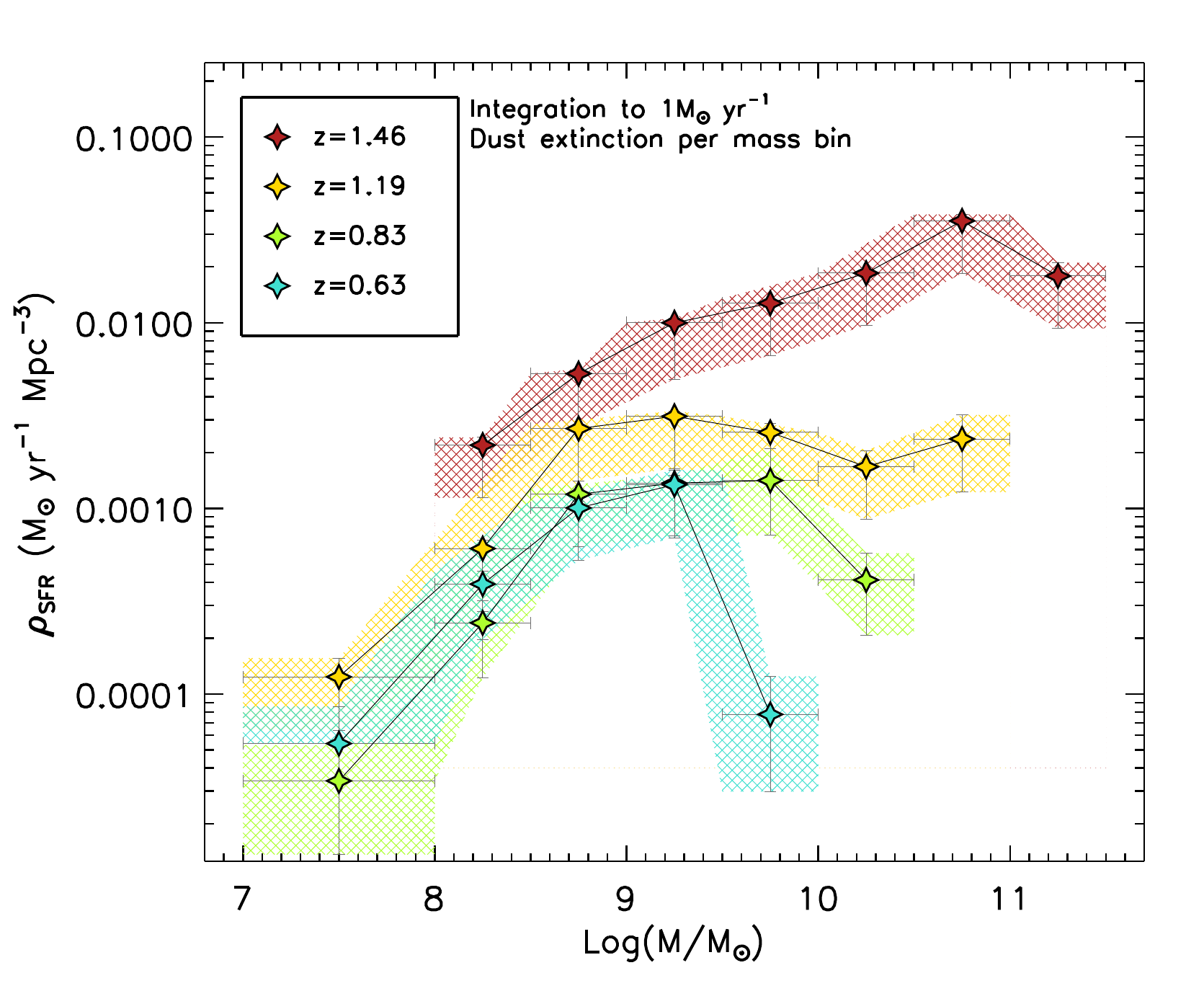}} 
\resizebox{0.48\textwidth}{!}{\includegraphics{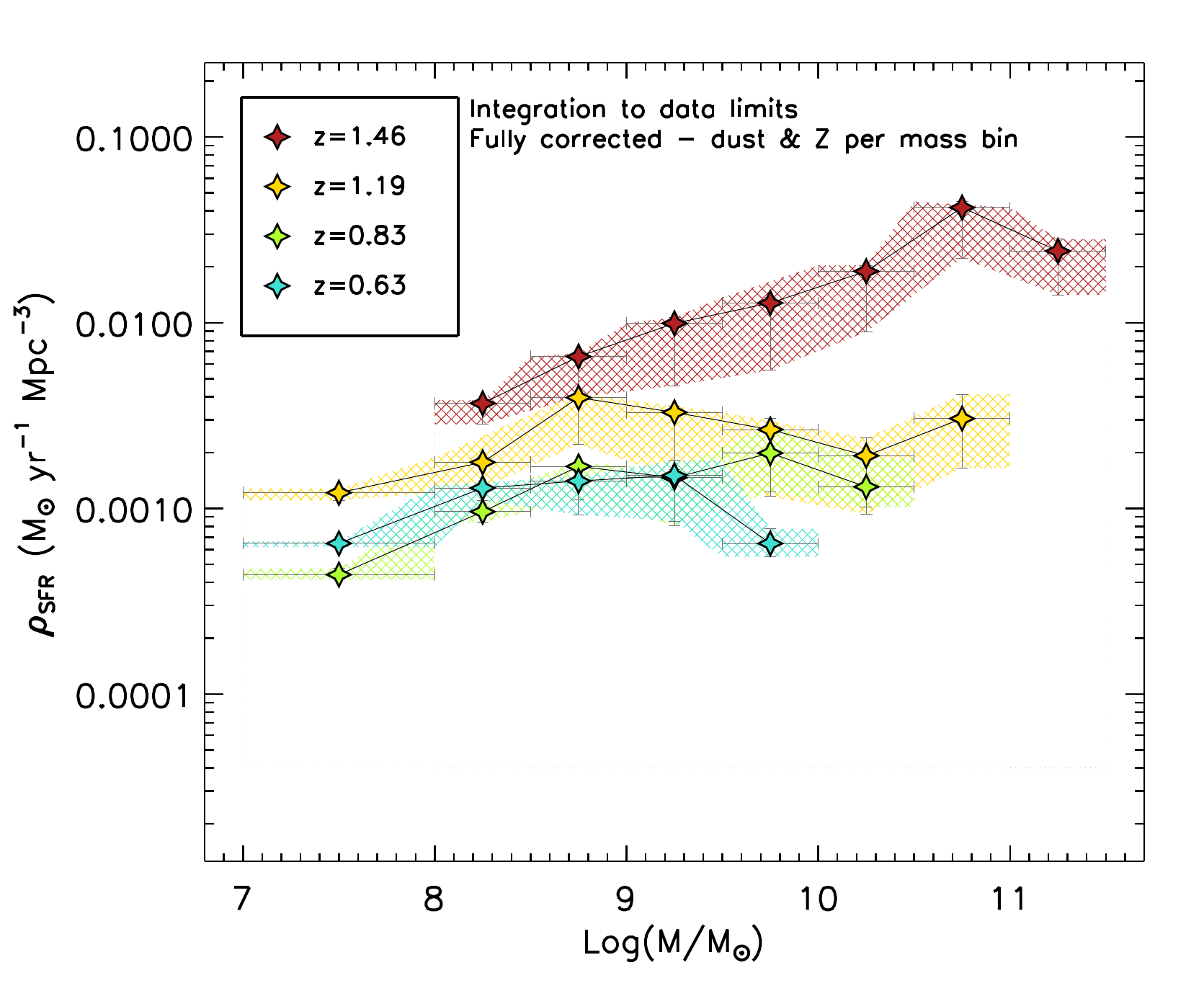}} 
\resizebox{0.48\textwidth}{!}{\includegraphics{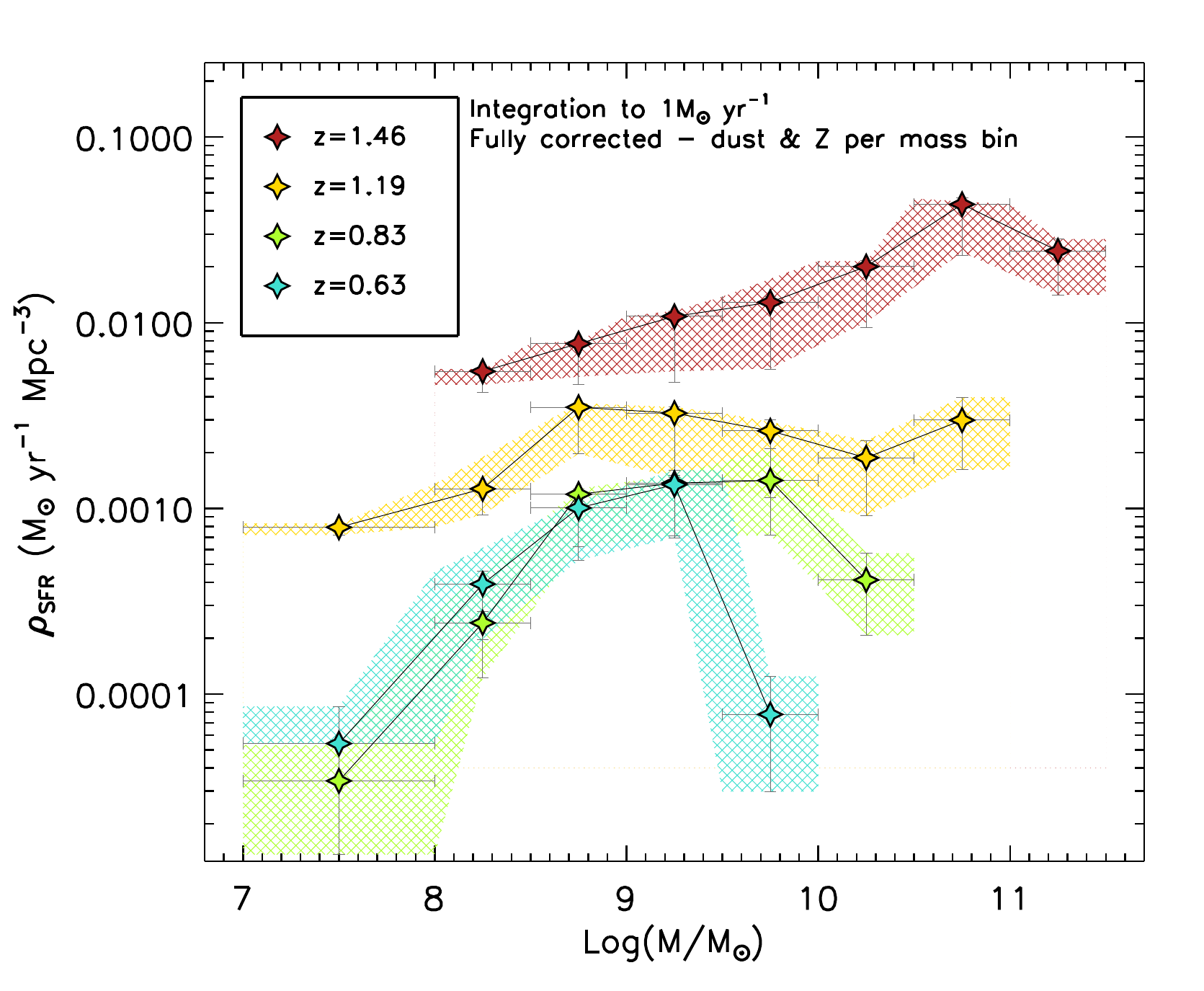}} 
\caption[]{$\rho_{\rm{SFR}}$ as a function of mass for redshift slices at $z=0.63$, $z=0.83$, $z=1.19$
and $z=1.46$. $\rho_{\rm{SFR}}$ is estimated via
  Kennicutt (1998) and 1 mag extinction at H$\alpha$ (top two panels), Kewley et al. (2004) and dust extinction as a function of stellar mass
(central two panels), and the fully corrected
$\rho_{\rm{SFR}}$ including dust an metallicity corrections (lower two
panels). The left-hand column of figures present results for an
integration to the limit of the data in each redshift slice, and the
right hand column of figures presents an integration to 1M$_{\odot}$yr$^{-1}$.}
\label{fig:SFRD(mass)}
\end{center}
\end{figure*}

 Errors on $\rho_{\rm{SFR}}$ are first
computed according to the maximum and minimum values that arise from
the 1$\sigma$ deviation of $\phi^*$, L$^*$ and $\alpha$ when all three
parameters are allowed to vary in the maximum-likelihood analysis. Additionally we quantify the
uncertainty in $\rho_{\rm{SFR}}$ introduced through the small number of
objects making up some mass bins by incorporating an additional
Poissonian error. The initial 1$\sigma$ error is
combined with the fractional error on $\rho_{\rm{SFR}}$ according
to \cite{Gehrels86} given the
number of objects making up the luminosity function. Errors quoted in
Table \ref{tab:SFRD} and shown on Figure \ref{fig:SFRD(mass)}
represent the combination of these two errors in quadrature.

\section{Discussion}
\label{discussion}

\subsection{The Effect of Dust and Metallicity Corrections on $\rho_{\rm{SFR}}$(M)}
Applying 1 magnitude of extinction at H$\alpha$ (top two panels of
Figure \ref{fig:SFRD(mass)}) shows
the peak of $\rho_{\rm{SFR}}$(M) lying in the
$10^{9.0}$\textless$\rm{M_{\odot}}$\textless $10^{9.5}$ bin at $z=1.46$ and
$z=1.19$, shifting to $10^{8.5}$\textless$\rm{M_{\odot}}$\textless $10^{9.0}$ at $z=0.83$, and
$z=0.63$, a feature which is preserved for both limits of integration. The most notable difference between the two
approaches is the far lower values of $\rho_{\rm{SFR}}$(M) in the two
lower redshift slices due to the $1 \rm{M_{\odot}} yr^{-1}$ limit probing
little below L$^*$. 

Incorporating dust corrections as a function of stellar mass (central
two panels) produces a dramatic change in the shape of the
$\rho_{\rm{SFR}}$(M) for the redshift $z=1.46$ slice, raising the
high mass end (\textgreater $10^{10.0}$$\rm{M_{\odot}}$)
significantly and lowering values of $\rho_{\rm{SFR}}$(M) for all
masses \textless $10^{10.0}$$\rm{M_{\odot}}$. Similarly the high-mass portions of
$\rho_{\rm{SFR}}$(M) in the $z=1.19$ and $z=0.83$ redshift slices
are lifted, however in the $z=0.63$ slice, little change is seen due
to the lack of massive objects. 

The lower two panels of Figure \ref{fig:SFRD(mass)} present our best estimates of $\rho_{\rm{SFR}}$(M) incorporating dust and
metallicity corrections. Little change is seen across the central
region of the function ($10^{8.0}$\textless$\rm{M_{\odot}}$\textless
$10^{10.0}$) however a small rise is seen in masses \textgreater $10^{10.0}$$\rm{M_{\odot}}$, and a large increase below $10^{8.0}$$\rm{M_{\odot}}$. This low mass increase only affects the $z=1.19$ slice
since galaxies below $10^{8.0}$$\rm{M_{\odot}}$ do not enter the
$z=1.46$ selection, and the low mass objects at $z=0.63$ and $z=0.83$
have their SFRs computed via [O{\sc~iii}] luminosity which requires no
metallicity correction. 

Figure \ref{fig:SFRD_contributions} presents
in a simple format the way each correction applied to the $\rho_{\rm{SFR}}$(M) affects
the shape of the function. In each panel the black shaded area and dashed line highlight the shape of the
$\rho_{\rm{SFR}}$(M) using the traditional assumptions about
[O{\sc~ii}]/H$\alpha$ and the converison of \cite{Kennicutt98}. The
dashed orange line then gives the $\rho_{\rm{SFR}}$(M)
corrected only for dust extinction as a function of stellar mass, and
the red dashed line gives the final $\rho_{\rm{SFR}}$(M) corrected
for extinction and metallicity (where appropriate) as a function of stellar mass for each redshift
slice.

\subsection{Evolution of $\rho_{\rm{SFR}}$(M)}
\begin{figure} 
\begin{center}
\resizebox{0.48\textwidth}{!}{\includegraphics{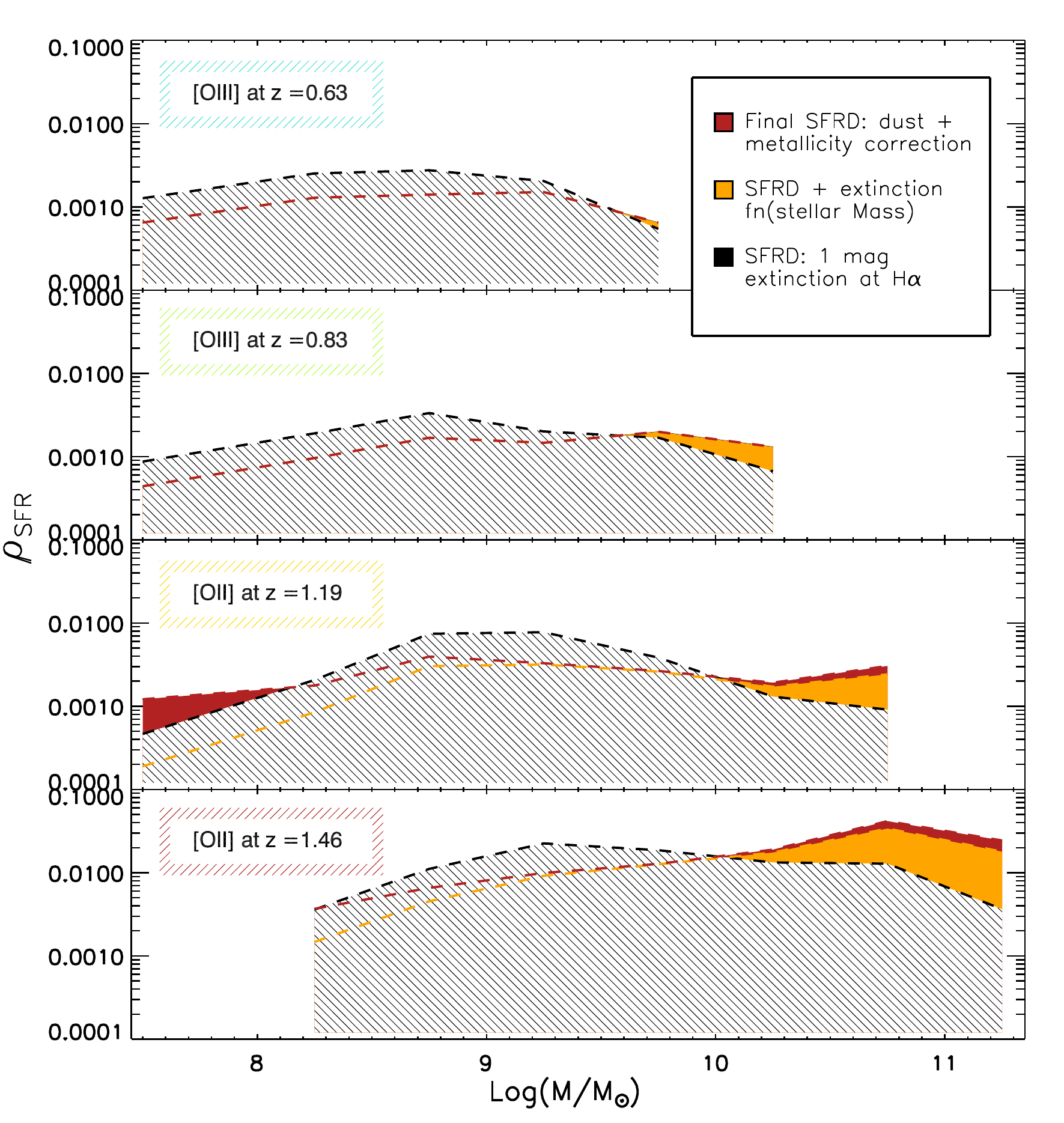}} 
\caption[]{The shape of the $\rho_{\rm{SFR}}$ as estimated via
  Kennicutt (1998) and 1 mag extinction at H$\alpha$ (black shaded region
plus black dashed line), Kewley et al. (2004) and dust extinction as a function of stellar mass
(orange shaded region and orange dashed line), and the fully corrected
$\rho_{\rm{SFR}}$ including dust an metallicity corrections (red
shaded region and red dashed line).}
\label{fig:SFRD_contributions}
\end{center}
\end{figure}

We favour the lower left-hand panel of Figure \ref{fig:SFRD(mass)} as
representative of the fully-corrected $\rho_{\rm{SFR}}$(M). The
peak contribution to $\rho_{\rm{SFR}}$ at $z=1.46$ comes
from galaxies in the mass range
$10^{10.5}$\textless$\rm{M_{\odot}}$\textless $10^{11.0}$, however in
the lower redshift slices it is less clear where the peak of the
function lies. The objects detected in the lower redshift slices do
not reach masses as high as those at $z=1.46$ and so it is impossible
to say if the $\rho_{\rm{SFR}}$ is still rising at these
masses. Figure \ref{fig:SFRD_contributions} highlights how dust
extinction as a function of mass kicks in at the highest masses which
would act to raise the $\rho_{\rm{SFR}}$ considerably at the high
masses which are missing from the lower two redshift slices.

The shape of the fully corrected $\rho_{\rm{SFR}}$(M) is remarkably
flat in comparison to the raw function, and shifts to lower normalisation with increasing time/decreasing
redshift. A similar effect is seen in both
\cite{Gilbank10} and \cite{Sobral14} who conclude that the shape of the $\rho_{\rm{SFR}}$ as a
function of mass shows very little evolution with redshift. One thing
we can note with this particular selection of galaxies is that
galaxies of very low mass ($10^{7.0}$\textless$\rm{M_{\odot}}$\textless $10^{8.0}$) play an important part in
the contribution to the overall $\rho_{\rm{SFR}}$ out to at least $z=1.19$.

\subsection{The Contribution of Different Masses to $\rho_{\rm{SFR}}$}
\begin{figure}
\begin{center}
\resizebox{0.48\textwidth}{!}{\includegraphics{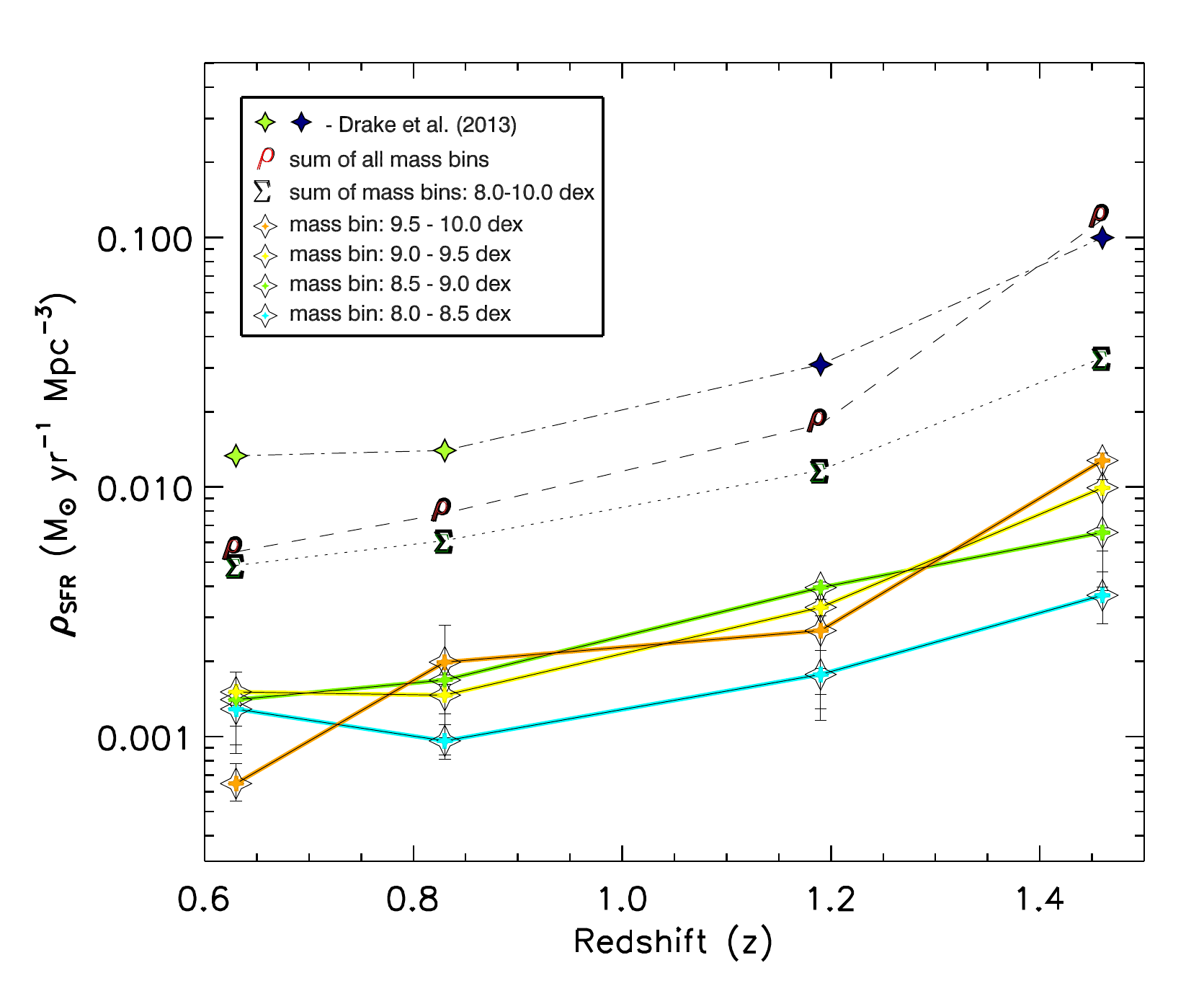}} 
\caption[]{Contribution of galaxies of different masses to the overall
  $\rho_{\rm{SFR}}$ in the four redshift slices studied
  here. Dot-dashed lines and points there-on represent the overall
  $\rho_{\rm{SFR}}$ taken from Drake et al (2013) converted to a
  Chabrier IMF. The dashed line and rho symbols present the sum of all
mass bins in this study, and the dotted line and capital Sigma symbols
show the sum of the four mass bins examined
here. $10^{8.0}$\textless$\rm{M_{\odot}}$\textless $10^{8.5}$ (cyan),
$10^{8.5}$\textless$\rm{M_{\odot}}$\textless $10^{9.0}$ (green), $10^{9.0}$\textless$\rm{M_{\odot}}$\textless $10^{9.5}$ (yellow), $10^{9.5}$\textless$\rm{M_{\odot}}$\textless $10^{10.0}$ (orange)}
\label{fig:SFRD_fn_z}
\end{center}
\end{figure}

Considering galaxies
in four mass bins where we have data points for all four redshift
slices: $10^{8.0}$\textless$\rm{M_{\odot}}$\textless $10^{8.5}$ (cyan),
$10^{8.5}$\textless$\rm{M_{\odot}}$\textless $10^{9.0}$ (green),
$10^{9.0}$\textless$\rm{M_{\odot}}$\textless $10^{9.5}$ (yellow),
$10^{9.5}$\textless$\rm{M_{\odot}}$\textless $10^{10.0}$ (orange),
Figure \ref{fig:SFRD_fn_z} examines the contributions from galaxies of
different masses to the overall $\rho_{\rm{SFR}}$. The dotted line and green $\Sigma$ symbols represent the sum of these
four bins, and coloured points represent the
contributions to $\rho_{\rm{SFR}}$ from each of these mass
bins. Broadly speaking the relative contributions from the different mass bins
decline simultaneously with the value of $\rho_{\rm{SFR}}$ itself. A
notable feature however is the stronger decline in
contribution from galaxies in the
$10^{9.50}$\textless$\rm{M_{\odot}}$\textless $10^{10.0}$ bin relative
to the lower mass bins.  This is in contrast to the result of
\cite{Sobral14} where $\rho_{\rm{SFR}}$ of galaxies in all three mass bins presented in
their figure 8 decline contemporaneously with the overall
$\rho_{\rm{SFR}}$. There are likely a number of reasons which lead to
this discrepancy. \cite{Sobral14} benefits from a consistently selected sample of
H$\alpha$ emitters across redshifts $z=0.40$ - $z=2.23$ meaning SFR is
calculated homogeneously across the entire redshift range. In contrast
we have estimated $\rho_{\rm{SFR}}$(M) using [O{\sc~iii}] ($z=0.63,
z=0.83$) or [O{\sc~ii}] ($z=1.19, z=1.46$) emission, indicators which
are known to have a higher and less well understood contamination from
AGN (see Section \ref{AGN}).

Additionally, slightly
different (larger) mass bins are used for the \cite{Sobral14}
analysis, plus the larger range in redshift studied may allow the small fluctuations seen across the
four redshift slices studied here to be smoothed out.

\subsection{The Effect of AGN Contamination on $\rho_{\rm{SFR}}$}
\label{AGN}

We can estimate the level of AGN-contamination by quantifying
  X-ray sources in the field that are coincident with the positions of our
  emission-line-selected sample. Using the catalogue of \cite{Ueda08}
  and matching within a 3 arcsecond radius, we find small numbers
  of interlopers in each redshift slice: 7, 1, 0 and 9 objects are
  found in redshift slices $z=0.63$, $z=0.83$, $z=1.19$ and $z=1.46$
  respectively, equating to $1$ percent (to the nearest percentage
  point) in the lowest redshift slice, and less than this in all other bins.

We take the same approach as in \cite{Drake13} and choose not to actively correct for AGN
contamination, since those objects with AGN activity will undoubtedly be
associated with some star formation, thus leading to a
possible over correction. 

We should however consider the consequences
of possible AGN contamination. If, as many authors have suggested, AGN
contamination rises towards higher redshift (e.g. 10 percent at
$z<1$, 15 percent at $z>1$  \citealt{Garn10}, \citealt{Sobral13}) the effect on $\rho_{\rm{SFR}}$ would be to lower
estimates above $z=1$ by a larger amount than estimates at $z<1$,
effectively moving the four coloured contours closer together, meaning
there has been less evolution of the $\rho_{\rm{SFR}}$ than implied in
Figure \ref{fig:SFRD(mass)}. 

A second consideration is the likelihood of the higher mass bins being
more highly contaminated by AGN than the lower mass bins. To place
some constraint on whether this may be the case, we chose to compare
the ratio of
line flux with flux at restframe 2500 \AA. While the measurements show a
correlation, the highest mass objects show the largest
discrepancy, with line fluxes that would produce a larger SFR than
that from the restframe 2500 \AA\,
flux. This gives an indication that the highest mass bins could be
overestimating $\rho_{\rm{SFR}}$ effectively boosting the peak of
$\rho_{\rm{SFR}}$ at the high mass end. If this is the case then the
shape of the $\rho_{\rm{SFR}}$(M) is infact flatter than depicted
in Fig \ref{fig:SFRD(mass)}. The physical implications of this
scenario are that the decline of star formation activity is even less
driven by the switch off of star formation in massive galaxies, and governed by a
process (or processes) acting across the entire mass range.

\section{Conclusions}
\label{conclusions}
We have applied the method described in \cite{Drake13} to determine
mass-binned luminosity functions in four redshift slices between
$z=1.46$ and $z=0.63$. We have investigated the effect of
extinction and metallicity corrections as a function of stellar mass,
and examined the role of galaxies of different masses in their
contribution to the overall $\rho_{\rm{SFR}}$. Our main conclusions can
be broadly summarised as follows:

\begin{itemize}
\renewcommand{\labelitemi}{$\bullet$}
\item The correct use of dust and metallicity corrections as a
  function of stellar mass is essential to establishing the true
  shape of $\rho_{\rm{SFR}}$(M) and its evolution with redshift. 
\item The normalisation of the fully corrected $\rho_{\rm{SFR}}$(M) shifts to lower values with
  increasing cosmic time (decreasing redshift).
\item The peak in $\rho_{\rm{SFR}}$(M) is seen in the
  $10^{10.5}$\textless$\rm{M_{\odot}}$\textless $10^{11.0}$ mass bin
  at $z=1.46$. In the lower redshift slices the location of the peak
  is less certain.  
\item Low mass galaxies reaching $10^{7.0}$\textless$\rm{M_{\odot}}$\textless $10^{8.0}$ play an important part in the $\rho_{\rm{SFR}}$(M). 
\item The contribution to the overall $\rho_{\rm{SFR}}$ from galaxies
  across the $10^{8.0}$\textless$\rm{M_{\odot}}$\textless
  $10^{9.5}$ mass range is roughly constant  between \mbox{$z=1.46$} and
  $z=0.63$. The contribution from the $10^{9.5}$\textless$\rm{M_{\odot}}$\textless
  $10^{10.0}$ bin drops more noticeably. 

\end{itemize}

In conclusion our results paint a picture in which the decline of
cosmic star formation activity since $z\sim1$ is the result of the
decline in SFR across a broad range of masses, i.e. it can not be
attributed simply to the switch-off of the most massive galaxies. This
brings to the forefront the question of the physical processes
required to execute such a decline, and the relative importance of
the different quenching mechanisms at work in truncating star formation.
        
\section*{Acknowledgements}
We thank the referee for a thorough read of the manuscript and many
insightful suggestions. S.Y. acknowledges support from Japan Society for the Promotion of
Science (JSPS) through JSPS research fellowship for foreign
researchers. JSD acknowledges the support of the European
Research Council via the award of an Advanced Grant,
and the contribution of the EC FP7 SPACE project ASTRODEEP
(Ref.No: 312725).

\bibliographystyle{mnras}
\bibliography{PAPERS2}

\appendix
\section{Maximum Likelihood fits}

\begin{table*}
\caption[Luminosity functions in mass bins for four redshift
  slices.]{Luminosity functions in mass bins for four redshift
  slices. Values presented here are raw results from maximum
  likelihood fitting. These numbers have not been corrected for
  aperture effects or dust extinction.}
\label{tab:massLF}
\renewcommand{\arraystretch}{1.2}
\begin{center}
\begin{tabular}{lcccccc} \hline

\multicolumn{5}{c}{\bf{Redshift $z= 0.63$ Volume= $8.09$} $\times$ $10^4$ Mpc$^3$}  \\
\cline{1-5}
{\bf{Log mass bin}}  & {\bf{Objects}}  & 
{\bf{log $\phi^{*}_{\rm{{[OIII]}}}$}}  & 
{\bf{log L$^{*}_{\rm{{[OIII]}}}$}} & 
{\bf{$\alpha_{\rm{{[OIII]}}}$}} \\
& && (Watts) & \\
\hline 
  7.00 $<$ {\bf{7.50}} $<$ 8.00 & 272 &   $-$3.08$^{+       0.35}_{-       0.55}$ &     33.90$^{+       0.30}_{-       0.21}$ &     $-$1.66$^{+       0.40}_{-       0.35}$ \\
8.00 $<$ {\bf{8.25}} $<$ 8.50  & 380 &     $-$2.60$^{+       0.08}_{-       0.13}$ &     33.86$^{+       0.14}_{-       0.11}$ &     $-$0.51$^{+       0.37}_{-       0.33}$ \\
 8.50 $<$ {\bf{8.75}} $<$ 9.00 & 202 &    $-$2.86$^{+       0.04}_{-       0.07}$ &     34.05$^{+       0.13}_{-       0.12}$ &      0.20$^{+       0.60}_{-       0.46}$ \\
 9.00 $<$ {\bf{9.25}} $<$ 9.50 &  62 &     $-$3.63$^{+       0.22}_{-       0.37}$ &     34.78$^{+       0.34}_{-       0.27}$ &     $-$0.72$^{+       0.63}_{-       0.44}$ \\
 9.50 $<$ {\bf{9.75}} $<$ 10.00 &  12 & $-$13.52$^{+ \infty  }_{- \infty  }$ &
 37.86$^{+ \infty}_{- \infty }$ & $-$3.37$^{+ \infty }_{- \infty}$\\
 9.50 $<$ {\bf{9.75}} $<$ 10.00 $^{\dagger}$ & 12 &     $-$3.73$^{+       0.49}_{-       0.90}$ &     34.06$^{+       0.46}_{-       0.26}$ &     $-$1.81$^{+       0.56}_{-       0.53}$ \\
\hline
\multicolumn{5}{c}{\bf{Redshift $z= 0.83$ Volume= $12.35$} $\times$ $10^4$ Mpc$^3$}  \\
\cline{1-5}
{\bf{Log mass bin}} & {\bf{Objects}}  & 
{\bf{log $\phi^{*}_{\rm{{[OIII]}}}$}}  & 
{\bf{log L$^{*}_{\rm{{[OIII]}}}$}} & 
{\bf{$\alpha_{\rm{{[OIII]}}}$}} \\

& & &
(Watts) & \\
\hline 

 7.00 $<$ {\bf{7.50}} $<$ 8.00 & 115 &    $-$2.72$^{+       0.09}_{-       0.28}$ &     33.55$^{+       0.27}_{-       0.19}$ &     $-$0.39$^{+       1.12}_{-       0.99}$ \\
 8.00 $<$ {\bf{8.25}} $<$ 8.50 &    250 &     $-$2.62$^{+       0.05}_{-       0.11}$ &     33.68$^{+       0.15}_{-       0.13}$ &      0.07$^{+       0.71}_{-       0.61}$ \\
 8.50 $<$ {\bf{8.75}} $<$ 9.00 & 315 &     $-$2.77$^{+       0.04}_{-       0.10}$ &     34.00$^{+       0.12}_{-       0.11}$ &      0.33$^{+       0.60}_{-       0.48}$ \\
 9.00 $<$ {\bf{9.25}} $<$ 9.50 & 122 &    $-$3.97$^{+       0.55}_{-       1.03}$ &     33.84$^{+       0.17}_{-       0.16}$ &      2.73$^{+       1.67}_{-       1.25}$ \\
 9.50 $<$ {\bf{9.75}} $<$ 10.00&   47 & $-$5.34$^{+ \infty }_{- \infty}$ & 35.78$^{+ \infty }_{- \infty}$ & $-$1.89$^{+ \infty }_{- \infty}$ \\
 9.50 $<$ {\bf{9.75}} $<$ 10.00 $^{\dagger}$& 47 &    $-$4.82$^{+       0.84}_{-       1.46}$ &     35.45$^{+       1.19}_{-       0.59}$ &     $-$1.73$^{+       0.50}_{-       0.28}$ \\
 10.00 $<$ {\bf{10.25}} $<$ 10.50 &   12 & $-$3.68 $^{+ \infty }_{- \infty}$ &  34.24$^{+ \infty }_{- \infty}$ & $-$2.60$^{+ \infty }_{- \infty}$ \\
 10.00 $<$ {\bf{10.25}} $<$ 10.50 $^{\dagger}$&  12 &    $-$3.68$^{+       0.72}_{-       0.95}$ &     33.45$^{+       0.17}_{-       0.13}$ &      2.11$^{+       1.53}_{-       1.54}$ \\
\hline
\multicolumn{5}{c}{\bf{Redshift $z= 1.19$ Volume= $19.06$} $\times$ $10^4$ Mpc$^3$}  \\
\cline{1-5}
{\bf{Log mass bin}} & {\bf{Objects}}  & 
{\bf{log $\phi^{*}_{\rm{{[OII]}}}$}}  & 
{\bf{log L$^{*}_{\rm{{[OII]}}}$}} & 
{\bf{$\alpha_{\rm{{[OII]}}}$}} \\

& & &
(Watts) &\\
\hline 

 7.00 $<$ {\bf{7.50}} $<$ 8.00  & 24 &   $-$4.59$^{+       1.06}_{-       5.00}$ &     34.56$^{+       5.01}_{-       0.59}$ &     $-$2.20$^{+       1.40}_{-       0.88}$ \\
 8.00 $<$ {\bf{8.25}} $<$ 8.50 &   118 & $-$6.53$^{+ \infty}_{- \infty }$ &  35.69$^{+ \infty}_{- \infty }$ 
 & $-$2.89$^{+ \infty}_{- \infty }$ \\
 8.00 $<$ {\bf{8.25}} $<$ 8.50 $^{\dagger}$&  118 &   $-$3.01$^{+       0.15}_{-       0.20}$ &     34.10$^{+       0.13}_{-       0.11}$ &     $-$1.26$^{+       0.30}_{-       0.31}$ \\
 8.50 $<$ {\bf{8.75}} $<$ 9.00 &  364 &    $-$2.62$^{+       0.07}_{-       0.11}$ &     34.22$^{+       0.11}_{-       0.10}$ &     $-$0.61$^{+       0.35}_{-       0.31}$ \\
 9.00 $<$ {\bf{9.25}} $<$ 9.50 &   259 &   $-$3.04$^{+       0.16}_{-       0.22}$ &     34.04$^{+       0.08}_{-       0.07}$ &      1.51$^{+       0.56}_{-       0.51}$ \\
 9.50 $<$ {\bf{9.75}} $<$ 10.00 & 103 &   $-$3.18$^{+       0.06}_{-       0.10}$ &     34.33$^{+       0.14}_{-       0.12}$ &      0.38$^{+       0.59}_{-       0.52}$ \\
 10.00 $<$ {\bf{10.25}} $<$ 10.50 & 33 &     $-$3.61$^{+       0.09}_{-       0.21}$ &     34.36$^{+       0.29}_{-       0.22}$ &      0.07$^{+       1.17}_{-       0.90}$ \\
 10.50 $<$ {\bf{10.75}} $<$ 11.00 & 16 &    $-$3.87$^{+       0.15}_{-       0.49}$ &     34.49$^{+       0.48}_{-       0.32}$ &     $-$0.12$^{+       1.81}_{-       1.21}$ \\

\hline
\multicolumn{5}{c}{\bf{Redshift $z= 1.46$ Volume= $23.06$} $\times$ $10^4$ Mpc$^3$}  \\
\cline{1-5}
{\bf{Log mass bin}} & {\bf{Objects}}  & 
{\bf{log $\phi^{*}_{\rm{{[OII]}}}$}}  & 
{\bf{log L$^{*}_{\rm{{[OII]}}}$}} & 
{\bf{$\alpha_{\rm{{[OII]}}}$}} \\

& & &
(Watts) & \\
\hline 
 8.00 $<$ {\bf{8.25}} $<$ 8.50 &   154 & $-$13.28$^{+ \infty}_{- \infty }$ & 37.73$^{+ \infty}_{- \infty }$
 & $-$3.88$^{+ \infty}_{- \infty }$ \\
8.00 $<$ {\bf{8.25}} $<$ 8.50 $^{\dagger}$& 154 &  $-$2.71$^{+       0.27}_{-       0.43}$ &     34.23$^{+       0.18}_{-       0.14}$ &     $-$2.09$^{+       0.46}_{-       0.44}$ \\
 8.50 $<$ {\bf{8.75}} $<$ 9.00 &  457 &   $-$2.22$^{+       0.05}_{-       0.10}$ &     34.14$^{+       0.11}_{-       0.10}$ &     $-$0.66$^{+       0.53}_{-       0.47}$ \\
 9.00 $<$ {\bf{9.25}} $<$ 9.50 & 637 &     $-$2.89$^{+       0.67}_{-       0.04}$ &     34.86$^{+       0.05}_{-       0.92}$ &     $-$1.43$^{+       4.62}_{-       0.02}$ \\
 9.50 $<$ {\bf{9.75}} $<$ 10.00 & 386 &    $-$2.68$^{+       0.05}_{-       0.08}$ &     34.44$^{+       0.08}_{-       0.07}$ &      0.56$^{+       0.39}_{-       0.35}$ \\
 10.00 $<$ {\bf{10.25}} $<$ 10.50 &  287 &  $-$2.99$^{+       0.14}_{-       0.20}$ &     34.80$^{+       0.14}_{-       0.11}$ &     $-$1.12$^{+       0.27}_{-       0.26}$ \\
 10.50 $<$ {\bf{10.75}} $<$ 11.00 & 219 &  $-$3.04$^{+       0.13}_{-       0.18}$ &     34.84$^{+       0.14}_{-       0.12}$ &     $-$0.97$^{+       0.31}_{-       0.28}$ \\
 11.00 $<$ {\bf{11.25}} $<$ 11.50 &  46 &    $-$4.17$^{+       0.58}_{-       1.30}$ &     34.35$^{+       0.19}_{-       0.18}$ &      2.41$^{+       2.11}_{-       1.47}$ \\
\hline
\hline
\end{tabular}
\end{center}
\end{table*}

\section{Values of $\rho_{\rm{SFR}}$(M)}

\begin{table*}
\caption[]{Values of $\rho_{\rm{SFR}}$ in mass bins for the four
  redshift slices studied here. Values of $\rho_{\rm{SFR}}$ are
  presented first with 1 magnitude of extinction at H$\alpha$, then
incorporating a treatment of extinction as a function of stellar mass,
and finally our best estimates of $\rho_{\rm{SFR}}$ incorporating both
dust and metallicity corrections (where appropriate) as a function of stellar mass.}
\label{tab:SFRD}
\renewcommand{\arraystretch}{1.4}
\begin{center}
\begin{tabular}{lcccccccc} \hline

\multicolumn{7}{c}{\bf{Redshift $z= 0.63$ Volume= $8.09$} $\times$ $10^4$ Mpc$^3$}  \\
\cline{1-7}
{\bf{Log mass}} &\multicolumn{2}{c}{1 mag H$\alpha$ \& Kennicutt 98}
&\multicolumn{2}{c}{Extinction(mass) \& Kewley et al. 04}  &
\multicolumn{2}{c}{Fully corrected} \\

(bin centre)&\multicolumn{2}{c}
{{\bf{$\rho_{\rm{SFR}_{[OIII]}}$}} (M$_{\odot}$ yr$^{-1}$ Mpc$^{-3}$)} &\multicolumn{2}{c}
{{\bf{$\rho_{\rm{SFR}_{[OIII]}}$}} (M$_{\odot}$ yr$^{-1}$ Mpc$^{-3}$)} &\multicolumn{2}{c}
{{\bf{$\rho_{\rm{SFR}_{[OIII]}}$}} (M$_{\odot}$ yr$^{-1}$ Mpc$^{-3}$)}\\
\cline{1-7}

\vspace{1mm}
&(\textgreater $0.08$)  & (\textgreater $1.00$)& (\textgreater
$0.08$)&(\textgreater $1.00$) & (\textgreater $0.08$) & (\textgreater $1.00$)  \\
\hline 

{\bf{7.50}} &     1.27E-03$^{+{\rm{  3.11E-05}}}_{-{\rm{  5.63E-05}}}$ &  1.06E-04$^{+{\rm{  6.22E-05}}}_{-{\rm{  6.56E-05}}}$
&   6.48E-04$^{+{\rm{  1.59E-05}}}_{-{\rm{  2.87E-05}}}$ &  5.41E-05$^{+{\rm{  3.17E-05}}}_{-{\rm{  0.00E+00}}}$
&    6.48E-04$^{+{\rm{  1.59E-05}}}_{-{\rm{  2.87E-05}}}$ &
5.41E-05$^{+{\rm{  3.17E-05}}}_{-{\rm{  0.00E+00}}}$ \\

{\bf{8.25}} &    2.52E-03$^{+{\rm{  8.39E-05}}}_{-{\rm{  3.71E-04}}}$
&  7.66E-04$^{+{\rm{  1.33E-04}}}_{-{\rm{  3.80E-04}}}$ &
1.29E-03$^{+{\rm{  4.28E-05}}}_{-{\rm{  1.89E-04}}}$ &
3.91E-04$^{+{\rm{  6.78E-05}}}_{-{\rm{  1.94E-04}}}$ &    1.29E-03$^{+{\rm{  4.28E-05}}}_{-{\rm{  1.89E-04}}}$ &  3.91E-04$^{+{\rm{  6.78E-05}}}_{-{\rm{  1.94E-04}}}$  \\

{\bf{8.75}} &   2.75E-03$^{+{\rm{  1.62E-04}}}_{-{\rm{  9.42E-04}}}$ &
1.97E-03$^{+{\rm{  1.76E-04}}}_{-{\rm{  9.43E-04}}}$ &
1.40E-03$^{+{\rm{  8.26E-05}}}_{-{\rm{  4.81E-04}}}$ &
1.01E-03$^{+{\rm{  8.96E-05}}}_{-{\rm{  4.81E-04}}}$ &    1.40E-03$^{+{\rm{  8.26E-05}}}_{-{\rm{  4.81E-04}}}$ &  1.01E-03$^{+{\rm{  8.96E-05}}}_{-{\rm{  4.81E-04}}}$ \\

{\bf{9.25}} &  2.04E-03$^{+{\rm{  4.11E-04}}}_{-{\rm{8.87E-04}}}$ &
1.82E-03$^{+{\rm{  3.63E-04}}}_{-{\rm{8.77E-04}}}$ &
1.51E-03$^{+{\rm{  3.03E-04}}}_{-{\rm{  6.54E-04}}}$ &
1.34E-03$^{+{\rm{  2.67E-04}}}_{-{\rm{  6.46E-04}}}$ &
1.51E-03$^{+{\rm{  3.03E-04}}}_{-{\rm{  6.54E-04}}}$ &
1.34E-03$^{+{\rm{  2.67E-04}}}_{-{\rm{  6.46E-04}}}$ \\

{\bf{9.75}} &    5.44E-04$^{+{\rm{  1.08E-04}}}_{-{\rm{  8.22E-05}}}$
&  6.53E-05$^{+{\rm{  3.93E-05}}}_{-{\rm{  4.02E-05}}}$ &
6.47E-04$^{+{\rm{  1.29E-04}}}_{-{\rm{  9.79E-05}}}$ &
7.77E-05$^{+{\rm{  4.68E-05}}}_{-{\rm{  4.78E-05}}}$ &  6.47E-04$^{+{\rm{  1.29E-04}}}_{-{\rm{  9.79E-05}}}$ &  7.77E-05$^{+{\rm{  4.68E-05}}}_{-{\rm{  4.78E-05}}}$ \\

\hline
\multicolumn{7}{c}{\bf{Redshift $z= 0.83$ Volume= $12.35$} $\times$ $10^4$ Mpc$^3$}  \\
\cline{1-7}
{\bf{Log mass}} &\multicolumn{2}{c}{1 mag H$\alpha$ \& Kennicutt 98}
&\multicolumn{2}{c}{Extinction(mass) \& Kewley et al. 04}  &
\multicolumn{2}{c}{Fully corrected} \\

(bin centre)&\multicolumn{2}{c}
{{\bf{$\rho_{\rm{SFR}_{[OIII]}}$}} (M$_{\odot}$ yr$^{-1}$ Mpc$^{-3}$)} &\multicolumn{2}{c}
{{\bf{$\rho_{\rm{SFR}_{[OIII]}}$}} (M$_{\odot}$ yr$^{-1}$ Mpc$^{-3}$)} &\multicolumn{2}{c}
{{\bf{$\rho_{\rm{SFR}_{[OIII]}}$}} (M$_{\odot}$ yr$^{-1}$ Mpc$^{-3}$)}\\
\cline{1-7}

\vspace{1mm}
&(\textgreater $0.12$)  & (\textgreater $1.00$)& (\textgreater
$0.12$)&(\textgreater $1.00$) & (\textgreater $0.12$) & (\textgreater $1.00$)  \\
\hline 
{\bf{7.50}} &    8.71E-04$^{+{\rm{  1.11E-04}}}_{-{\rm{  4.17E-04}}}$
&  6.74E-05$^{+{\rm{  3.78E-05}}}_{-{\rm{  4.04E-05}}}$ &
4.40E-04$^{+{\rm{  3.39E-05}}}_{-{\rm{  2.53E-05}}}$ &
3.41E-05$^{+{\rm{  1.91E-05}}}_{-{\rm{  2.04E-05}}}$ &
4.40E-04$^{+{\rm{  3.39E-05}}}_{-{\rm{  2.53E-05}}}$ &
3.41E-05$^{+{\rm{  1.91E-05}}}_{-{\rm{  2.04E-05}}}$ \\

{\bf{8.25}} &    1.90E-03$^{+{\rm{  1.39E-04}}}_{-{\rm{  9.08E-04}}}$
&  4.78E-04$^{+{\rm{  7.15E-05}}}_{-{\rm{  2.36E-04}}}$ &
9.60E-04$^{+{\rm{  3.18E-05}}}_{-{\rm{  1.19E-04}}}$ &
2.42E-04$^{+{\rm{  3.61E-05}}}_{-{\rm{  1.19E-04}}}$ &
9.60E-04$^{+{\rm{  3.18E-05}}}_{-{\rm{  1.19E-04}}}$ &
2.42E-04$^{+{\rm{  3.61E-05}}}_{-{\rm{  1.19E-04}}}$   \\

{\bf{8.75}} &   3.33E-03$^{+{\rm{  2.21E-04}}}_{-{\rm{  1.59E-03}}}$ &
2.36E-03$^{+{\rm{  1.68E-04}}}_{-{\rm{  1.13E-03}}}$ &
1.68E-03$^{+{\rm{  8.63E-05}}}_{-{\rm{  5.70E-04}}}$ &
1.19E-03$^{+{\rm{  8.50E-05}}}_{-{\rm{  5.70E-04}}}$ &
1.68E-03$^{+{\rm{  8.63E-05}}}_{-{\rm{  5.70E-04}}}$ &
1.19E-03$^{+{\rm{  8.50E-05}}}_{-{\rm{  5.70E-04}}}$    \\

{\bf{9.25}} &    2.00E-03$^{+{\rm{  2.23E-04}}}_{-{\rm{  9.58E-04}}}$
&  1.87E-03$^{+{\rm{  1.99E-04}}}_{-{\rm{  8.94E-04}}}$ &
1.46E-03$^{+{\rm{  1.55E-04}}}_{-{\rm{  6.54E-04}}}$ &
1.37E-03$^{+{\rm{  1.45E-04}}}_{-{\rm{  6.53E-04}}}$ &
1.46E-03$^{+{\rm{  1.55E-04}}}_{-{\rm{  6.54E-04}}}$ &
1.37E-03$^{+{\rm{  1.45E-04}}}_{-{\rm{  6.53E-04}}}$  \\

{\bf{9.75}} &   1.68E-03$^{+{\rm{  7.14E-04}}}_{-{\rm{  8.55E-04}}}$ &
1.20E-03$^{+{\rm{  5.75E-04}}}_{-{\rm{  5.92E-04}}}$ &
1.98E-03$^{+{\rm{  8.08E-04}}}_{-{\rm{  7.57E-04}}}$ &
1.41E-03$^{+{\rm{  6.78E-04}}}_{-{\rm{  6.98E-04}}}$ &
1.98E-03$^{+{\rm{  8.08E-04}}}_{-{\rm{  7.57E-04}}}$ &
1.41E-03$^{+{\rm{  6.78E-04}}}_{-{\rm{  6.98E-04}}}$  \\

{\bf{10.25}} &   6.68E-04$^{+{\rm{  3.72E-04}}}_{-{\rm{  3.37E-04}}}$
&  2.10E-04$^{+{\rm{  8.26E-05}}}_{-{\rm{  1.04E-04}}}$ &
1.31E-03$^{+{\rm{  5.53E-04}}}_{-{\rm{  2.91E-04}}}$ &
4.11E-04$^{+{\rm{  1.62E-04}}}_{-{\rm{  2.04E-04}}}$ &
1.31E-03$^{+{\rm{  5.53E-04}}}_{-{\rm{  2.91E-04}}}$ &
4.11E-04$^{+{\rm{  1.62E-04}}}_{-{\rm{  2.04E-04}}}$  \\

\hline
\multicolumn{7}{c}{\bf{Redshift $z= 1.19$ Volume= $19.06$} $\times$ $10^4$ Mpc$^3$}  \\
\cline{1-7}
{\bf{Log mass}} &\multicolumn{2}{c}{1 mag H$\alpha$ \& Kennicutt 98}
&\multicolumn{2}{c}{Extinction(mass) \& Kewley et al. 04}  &
\multicolumn{2}{c}{Fully corrected: Extinction(mass) + Z} \\

(bin centre)&\multicolumn{2}{c}
{{\bf{$\rho_{\rm{SFR}_{[OII]}}$}} (M$_{\odot}$ yr$^{-1}$ Mpc$^{-3}$)} &\multicolumn{2}{c}
{{\bf{$\rho_{\rm{SFR}_{[OII]}}$}} (M$_{\odot}$ yr$^{-1}$ Mpc$^{-3}$)} &\multicolumn{2}{c}
{{\bf{$\rho_{\rm{SFR}_{[OII]}}$}} (M$_{\odot}$ yr$^{-1}$ Mpc$^{-3}$)}\\
\cline{1-7}

\vspace{1mm}
&(\textgreater $0.55$)  & (\textgreater $1.00$)& (\textgreater
$0.55$)&(\textgreater $1.00$) & (\textgreater $0.55$) & (\textgreater $1.00$)  \\
\hline 
{\bf{7.50}} &   4.65E-04$^{+{\rm{  1.33E-04}}}_{-{\rm{  2.28E-04}}}$ &
3.03E-04$^{+{\rm{  7.94E-05}}}_{-{\rm{  1.47E-04}}}$ &
1.90E-04$^{+{\rm{  5.43E-05}}}_{-{\rm{  9.32E-05}}}$ &
1.24E-04$^{+{\rm{  3.25E-05}}}_{-{\rm{  6.02E-05}}}$ &
1.21E-03$^{+{\rm{  6.52E-05}}}_{-{\rm{  1.12E-04}}}$ &
7.91E-04$^{+{\rm{  3.90E-05}}}_{-{\rm{  7.23E-05}}}$  \\

{\bf{8.25}} &    2.07E-03$^{+{\rm{  2.18E-04}}}_{-{\rm{  9.87E-04}}}$
&  1.49E-03$^{+{\rm{  1.59E-04}}}_{-{\rm{  7.11E-04}}}$ &
8.45E-04$^{+{\rm{  8.90E-05}}}_{-{\rm{  4.04E-04}}}$ &
6.08E-04$^{+{\rm{  6.49E-05}}}_{-{\rm{  2.90E-04}}}$ &
1.77E-03$^{+{\rm{  1.07E-04}}}_{-{\rm{  4.84E-04}}}$ &
1.27E-03$^{+{\rm{  7.80E-05}}}_{-{\rm{  3.49E-04}}}$ \\

{\bf{8.75}} &   7.42E-03$^{+{\rm{  4.44E-04}}}_{-{\rm{  3.55E-03}}}$ &
6.59E-03$^{+{\rm{  3.71E-04}}}_{-{\rm{  3.15E-03}}}$ &
3.03E-03$^{+{\rm{  1.82E-04}}}_{-{\rm{  1.45E-03}}}$ &
2.69E-03$^{+{\rm{  1.52E-04}}}_{-{\rm{  1.29E-03}}}$ &
3.95E-03$^{+{\rm{  2.18E-04}}}_{-{\rm{  1.74E-03}}}$ &
3.51E-03$^{+{\rm{  1.82E-04}}}_{-{\rm{  1.54E-03}}}$  \\

{\bf{9.25}} &    7.75E-03$^{+{\rm{  5.35E-04}}}_{-{\rm{  3.70E-03}}}$
&  7.67E-03$^{+{\rm{  5.20E-04}}}_{-{\rm{  3.66E-03}}}$ &
3.17E-03$^{+{\rm{  2.19E-04}}}_{-{\rm{  1.51E-03}}}$ &
3.13E-03$^{+{\rm{  2.12E-04}}}_{-{\rm{  1.50E-03}}}$ &
3.29E-03$^{+{\rm{  2.62E-04}}}_{-{\rm{  1.81E-03}}}$ &
3.26E-03$^{+{\rm{  2.55E-04}}}_{-{\rm{  1.80E-03}}}$ \\

{\bf{9.75}} &    3.84E-03$^{+{\rm{  4.80E-04}}}_{-{\rm{  1.84E-03}}}$
&  3.78E-03$^{+{\rm{  4.63E-04}}}_{-{\rm{  1.81E-03}}}$ &
2.61E-03$^{+{\rm{  3.25E-04}}}_{-{\rm{  1.25E-03}}}$ &
2.57E-03$^{+{\rm{  3.14E-04}}}_{-{\rm{  1.23E-03}}}$ &
2.66E-03$^{+{\rm{  3.90E-04}}}_{-{\rm{  1.50E-03}}}$ &
2.62E-03$^{+{\rm{  3.77E-04}}}_{-{\rm{  1.47E-03}}}$  \\

{\bf{10.25}} &    1.31E-03$^{+{\rm{  2.96E-04}}}_{-{\rm{  6.27E-04}}}$
&  1.27E-03$^{+{\rm{  2.85E-04}}}_{-{\rm{  6.10E-04}}}$ &
1.72E-03$^{+{\rm{  3.90E-04}}}_{-{\rm{  8.25E-04}}}$ &
1.68E-03$^{+{\rm{  3.75E-04}}}_{-{\rm{  8.03E-04}}}$ &
1.92E-03$^{+{\rm{  4.68E-04}}}_{-{\rm{  9.91E-04}}}$ &
1.87E-03$^{+{\rm{  4.50E-04}}}_{-{\rm{  9.64E-04}}}$ \\

{\bf{10.75}} &    9.11E-04$^{+{\rm{  3.35E-04}}}_{-{\rm{  4.42E-04}}}$
&  8.92E-04$^{+{\rm{  3.08E-04}}}_{-{\rm{  4.31E-04}}}$ &
2.41E-03$^{+{\rm{  8.89E-04}}}_{-{\rm{  1.17E-03}}}$ &
2.37E-03$^{+{\rm{  8.18E-04}}}_{-{\rm{  1.14E-03}}}$ &
3.05E-03$^{+{\rm{  1.07E-03}}}_{-{\rm{  1.41E-03}}}$ &
2.99E-03$^{+{\rm{  9.82E-04}}}_{-{\rm{  1.37E-03}}}$  \\

\hline
\multicolumn{7}{c}{\bf{Redshift $z= 1.46$ Volume= $23.06$} $\times$ $10^4$ Mpc$^3$}  \\
\cline{1-7}
{\bf{Log mass}} &\multicolumn{2}{c}{1 mag H$\alpha$ \& Kennicutt 98}
&\multicolumn{2}{c}{Extinction(mass) \& Kewley et al. 04}  &
\multicolumn{2}{c}{Fully corrected: Extinction(mass) + Z} \\

(bin centre)&\multicolumn{2}{c}
{{\bf{$\rho_{\rm{SFR}_{[OII]}}$}} (M$_{\odot}$ yr$^{-1}$ Mpc$^{-3}$)} &\multicolumn{2}{c}
{{\bf{$\rho_{\rm{SFR}_{[OII]}}$}} (M$_{\odot}$ yr$^{-1}$ Mpc$^{-3}$)} &\multicolumn{2}{c}
{{\bf{$\rho_{\rm{SFR}_{[OII]}}$}} (M$_{\odot}$ yr$^{-1}$ Mpc$^{-3}$)}\\
\cline{1-7}

\vspace{1mm}
&(\textgreater $1.42$)  & (\textgreater $1.00$)& (\textgreater
$1.42$)&(\textgreater $1.00$) & (\textgreater $1.42$) & (\textgreater $1.00$)  \\
\hline 
{\bf{8.25}} &   3.61E-03$^{+{\rm{  3.18E-04}}}_{-{\rm{  1.72E-03}}}$ &
5.37E-03$^{+{\rm{  5.42E-04}}}_{-{\rm{  2.58E-03}}}$ &
1.47E-03$^{+{\rm{  1.30E-04}}}_{-{\rm{  7.03E-04}}}$ &
2.19E-03$^{+{\rm{  2.21E-04}}}_{-{\rm{  1.05E-03}}}$ &
3.68E-03$^{+{\rm{  1.56E-04}}}_{-{\rm{  8.44E-04}}}$ &
5.48E-03$^{+{\rm{  2.65E-04}}}_{-{\rm{  1.26E-03}}}$ \\

{\bf{8.75}} &    1.11E-02$^{+{\rm{  5.54E-04}}}_{-{\rm{  5.30E-03}}}$
&  1.30E-02$^{+{\rm{  7.91E-04}}}_{-{\rm{  6.24E-03}}}$ &
4.53E-03$^{+{\rm{  2.26E-04}}}_{-{\rm{  2.16E-03}}}$ &
5.32E-03$^{+{\rm{  3.23E-04}}}_{-{\rm{  2.55E-03}}}$ &
6.57E-03$^{+{\rm{  2.71E-04}}}_{-{\rm{  2.60E-03}}}$ &
7.71E-03$^{+{\rm{  3.88E-04}}}_{-{\rm{  3.06E-03}}}$  \\

{\bf{9.25}} &     2.25E-02$^{+{\rm{  1.62E-03}}}_{-{\rm{  1.09E-02}}}$
&  2.45E-02$^{+{\rm{  1.46E-03}}}_{-{\rm{  1.23E-02}}}$ &
9.17E-03$^{+{\rm{  6.59E-04}}}_{-{\rm{  4.45E-03}}}$ &
1.00E-02$^{+{\rm{  5.96E-04}}}_{-{\rm{  5.03E-03}}}$ &
9.92E-03$^{+{\rm{  7.91E-04}}}_{-{\rm{  5.34E-03}}}$ &
1.08E-02$^{+{\rm{  7.16E-04}}}_{-{\rm{  6.04E-03}}}$ \\

{\bf{9.75}} &   1.86E-02$^{+{\rm{  1.11E-03}}}_{-{\rm{  8.90E-03}}}$ &
1.88E-02$^{+{\rm{  1.15E-03}}}_{-{\rm{  8.98E-03}}}$ &
1.26E-02$^{+{\rm{  7.54E-04}}}_{-{\rm{  6.03E-03}}}$ &
1.27E-02$^{+{\rm{  7.80E-04}}}_{-{\rm{  6.08E-03}}}$ &
1.28E-02$^{+{\rm{  9.06E-04}}}_{-{\rm{  7.23E-03}}}$ &
1.29E-02$^{+{\rm{  9.37E-04}}}_{-{\rm{  7.30E-03}}}$  \\

{\bf{10.25}} &   1.33E-02$^{+{\rm{  8.99E-04}}}_{-{\rm{  6.34E-03}}}$
&  1.41E-02$^{+{\rm{  1.02E-03}}}_{-{\rm{  6.73E-03}}}$ &
1.75E-02$^{+{\rm{  1.18E-03}}}_{-{\rm{  8.34E-03}}}$ &
1.85E-02$^{+{\rm{  1.33E-03}}}_{-{\rm{  8.84E-03}}}$ &
1.89E-02$^{+{\rm{  1.42E-03}}}_{-{\rm{  1.00E-02}}}$ &
2.01E-02$^{+{\rm{  1.60E-03}}}_{-{\rm{  1.06E-02}}}$ \\

{\bf{10.75}} &  1.28E-02$^{+{\rm{  1.01E-03}}}_{-{\rm{  6.13E-03}}}$ &
1.34E-02$^{+{\rm{  1.11E-03}}}_{-{\rm{  6.39E-03}}}$ &
3.40E-02$^{+{\rm{  2.68E-03}}}_{-{\rm{  1.62E-02}}}$ &
3.54E-02$^{+{\rm{  2.94E-03}}}_{-{\rm{  1.69E-02}}}$ &
4.17E-02$^{+{\rm{  3.22E-03}}}_{-{\rm{  1.95E-02}}}$ &
4.34E-02$^{+{\rm{  3.53E-03}}}_{-{\rm{  2.03E-02}}}$ \\

{\bf{11.25}} &    3.64E-03$^{+{\rm{  6.51E-04}}}_{-{\rm{  1.74E-03}}}$
&  3.64E-03$^{+{\rm{  6.54E-04}}}_{-{\rm{  1.74E-03}}}$ &
1.78E-02$^{+{\rm{  3.19E-03}}}_{-{\rm{  8.52E-03}}}$ &
1.78E-02$^{+{\rm{  3.21E-03}}}_{-{\rm{  8.53E-03}}}$ &
2.44E-02$^{+{\rm{  3.84E-03}}}_{-{\rm{  1.02E-02}}}$ &
2.44E-02$^{+{\rm{  3.85E-03}}}_{-{\rm{  1.02E-02}}}$  \\
\hline
\end{tabular}
\end{center}
\end{table*}

\label{lastpage}

\end{document}